\documentclass{article}

\usepackage{thumbpdf,lmodern}

\usepackage{framed}
\usepackage{appendix}
\usepackage{xcolor}

\RequirePackage{amsthm,amsmath,amsfonts,amssymb}
\RequirePackage[authoryear]{natbib}
\RequirePackage{graphicx, tikz, pgfplots}
\RequirePackage[ruled,vlined]{algorithm2e}
\RequirePackage{appendix, float, ulem, float, url}
\usepackage{natbib, ulem, authblk, float, setspace, booktabs}
\usepackage[margin=1in]{geometry}

\definecolor{pptblue}{RGB}{68, 114, 196}
\definecolor{pptred}{RGB}{210, 0, 0}

\SetKwInput{KwInputs}{Inputs}
\SetKwInput{KwOutput}{Output}

\newcommand{\fct}[1]{\texttt{#1()}}

\pgfplotsset{compat=1.17}

\title{\textbf{fabisearch}: A Package for Change Point Detection in and Visualization of the Network Structure of Multivariate High-Dimensional Time Series in R}

\author[1]{Martin Ondrus}
\author[1,2]{Ivor Cribben}
\affil[1]{Neuroscience and Mental Health Institute, University of Alberta}
\affil[2]{Alberta School of Business, University of Alberta}

\begin{document}

\maketitle{}

\begin{abstract}
In this work, we introduce the R package \textbf{fabisearch}, available on the Comprehensive R Archive Network (CRAN), which implements an original change point detection method for multivariate high-dimensional time series data and a new interactive, 3-dimensional, brain-specific network visualization capability in a flexible, stand-alone function.  Change point detection is a commonly used technique in time series analysis, capturing the dynamic nature in which many real-world processes function. With the ever increasing troves of multivariate high-dimensional time series data, especially in neuroimaging and finance, there is a clear need for scalable and data-driven change point detection methods. Currently, change point detection methods for multivariate high-dimensional data are scarce, with even less available in high-level, easily accessible software packages. \textbf{fabisearch}, which implements the \textit{factorized binary search} (FaBiSearch) methodology, is a novel statistical method for detecting change points in the network structure of multivariate high-dimensional time series which employs \textit{non-negative matrix factorization} (NMF), an unsupervised dimension reduction and clustering technique. Given the high computational cost of NMF, we implement the method in C++ code and use parallelization to reduce computation time. Further, we also utilize a new binary search algorithm to efficiently identify multiple change points and provide a new method for network estimation for data between change points. We show the functionality of the package and the practicality of the method by applying it to a neuroimaging and a finance data set. We also introduce an interactive, 3-dimensional, brain-specific network visualization capability in a flexible, stand-alone function. This function can be conveniently used with any node coordinate atlas, and nodes can be color coded according to community membership (if applicable). The output is an elegantly displayed network laid over a cortical surface, which can be rotated in the 3-dimensional space.
\end{abstract}

{\bf Keywords: change point detection, time series, high-dimensional, NMF, fMRI, network analysis, R, visualization, bioinformatics}

\newpage 

\doublespacing

\section{Introduction} \label{sec:intro}

Change point detection is a collection of methods in time series analysis that identify if and when changes occur in the statistical properties of time series data. Consequently, it is a useful way of studying the evolution of the underlying data generating process. In practice, changes might correspond to changes in the mean, the variance, or, for multivariate time series data, changes in the covariance between the marginal time series. Change points convey useful information and can be applied to a wide variety of problems such as the active management of financial assets \citep{lai2015active}, anomaly detection in computer networks for cybersecurity \citep{Tartakovsky2013}, or investigating dynamic brain networks during task performance \citep{Cribben2012}.

Given the utility of change point detection, numerous methods have been developed.  For a thorough review on the available methods, see \cite{truong}.  Many R packages have also been made available. There are some packages available with a specific application in mind such as genetic/genomics data, for example, \textbf{DNAcopy} \citep{Seshan2020}, \textbf{Rseg} \citep{10.1093/bioinformatics/btq668}, and \textbf{cumseg} \citep{Muggeo2020}. Other packages exist for more general applications, such as the \textbf{changepoint} package \citep{JSSv058i03} which includes a variety of single and multiple change point detection methods. The \textbf{strucchange} package \citep{JSSv007i02} attempts to find structural breaks in univariate time series. However, it focuses on detecting at most one change point (AMOC) and is motivated by changes in (linear) regression models. The \textbf{cpm} package \citep{JSSv066i03} provides functionality to detect change points in univariate time series using a variety of different methods: from distribution-free change detection methods for changes in the mean, variance, or general distribution to parametric change detection methods for Gaussian, Bernoulli and Exponential sequences. The \textbf{mosum} package \citep{JSSv097i08} provides functionality for finding changes in the mean in a univariate time series using the moving sum statistic. For complex multivariate systems (such as neuroimaging data), however, these methods are unable to capture changes which occur from the interaction of multiple univariate components.

Although not as widely available, some R packages for change point detection in multivariate time series data do exist.  The \textbf{bcp} package \citep{JSSv023i03} was extended to multivariate problems through with the work of \cite{Wang2015}, who extended the methodology to find change points on general connected graphs using Bayesian methods. \cite{Chen2015} proposed a graph based change point detection method in the \textbf{gSeg}  package \citep{Chen2020}, but it requires a pre-defined similarity measure.  The \textbf{ecp} package \citep{JSSv097i08} attempts to find multiple change points while making minimal assumptions about the data generating process, and provides functionality for non-parametric change point detection for both univariate and multivariate time series data. \cite{Grundy2020} suggest a novel geometric method to reduce the change point problem to two dimensions and then apply change point detection methods to find changes in the mean and in the variance. They compare their new method to \textbf{ecp}, demonstrating its superior performance and making it available as the \textbf{changepoint.geo} package \citep{Grundy2020a}.  The \textbf{onlineCOV} \citep{Li2020a} package implements the work of \cite{Li2020} who propose an online change point detection method which can be applied to both Gaussian and non-Gaussian data. This method, however, requires training in order for the time series dependence structure to be learned. \cite{Londschien2020a} suggest a Gaussian graphical model based method for detecting change points in applications with missing data. These methods are available in the \textbf{hdcd} package available on GitHub (\url{https://github.com/mlondschien/hdcd}). \cite{Xiong2021} developed the Vine Copula Change Point (VCCP) method, which uses vine copulas, and thus can detect changes in dependence structures beyond the linear, Gaussian assumptions that dominate most models. This methodology is available in the \textbf{vccp} package \citep{Xiong2021a}. Lastly, \cite{anastasiou} developed the Cross-Covariance Isolate Detect (CCID) method, which considers changes in the cross-covariance structure in a multivariate (possibly high-dimensional) time series.  Using a wavelet based approach, it transforms the multivariate time series and then uses a CUSUM based statistic to detect change points. This method is available in the \textbf{ccid} package \citep{Anastasiou2020a}.

Neuroimaging data provides an especially unique opportunity to apply change point detection methods. Functional magnetic resonance imaging (fMRI) is a popular method of indirectly measuring brain activity using the blood-oxygen-level-dependent (BOLD) signal \citep{ogawa}. With increases in neuronal activity, the BOLD signal increases and shows brain regions ``lighting up'' in fMRI images. Multiple slices of each subject's brain are imaged and divided into voxels, which are small cubes of a few millimeters in dimension. Subsequently, a time series of voxel activity can be measured by taking multiple fMRI scans sequentially over time. It is of great interest to understand brain dynamics through these time series of voxel activity, or, more typically, the activity in a cluster of voxels referred to as a region of interest (ROI). Functional connectivity (FC) seeks to define relationships between the voxel or ROI time series \citep{biswal}, through the use of correlation, covariance, precision matrices, or other methods (see \citealp{cribbenfiecas} for a comprehensive review). Using graph theory, we can intuitively understand FC as a graphical model where nodes represent ROIs (time series) and edges represent temporal relationships. 

Although change point detection has been applied to multivariate fMRI time series data \citep{Cribben2012,barnett,dai2019,ofori2019,anastasiou}, most methods and software lack an ability to consider a multivariate high-dimensional time series or whole brain dynamics that are a result of applying high-dimensional cortical parcellations (e.g., the Gordon atlas with $p = 333$ ROIs, \citealp{Gordon2016}).  In addition, many of the existing R packages require \textit{a priori} knowledge or assumptions, which makes them less accessible to a general audience who may not have domain expertise in change point detection. Furthermore, many of the existing R packages do not provide functionality to both estimate and visualize (brain-specific) networks between change points. To this end, we introduce the R package, \textbf{fabisearch} \citep{ondrus_fabi}, which implements the factorized binary search (FaBiSearch) methodology \citep{ondrus2021factorized}. FaBiSearch is a powerful technique which employs non-negative matrix factorization (NMF) to detect changes in the network (or clustering) structure of multivariate high-dimensional time series data. By integrating NMF as the dependency modelling method in change point detection, FaBiSearch scales to data sets where the number of variables (or time seres) is in the 100s or 1000s, and in particular for the case where the number of time series is greater than the number of time points ($p >> T$). In addition, the NMF element of FaBiSearch allows us to define a new method for the estimation of networks for data between change points, which provides a visual display of the clustering structure and the networks between the variables (or time series). Our method also incorporates a novel binary search based algorithm for change point detection.

The second contribution in the \textbf{fabisearch} package is the functionality to estimate and visualize networks between change points. In particular, it provides a flexible stand-alone function that has the ability to visualize an interactive, 3-dimensional, brain-specific network. To the best of our knowledge, the only other R package which includes similar capability is \textbf{brainGraph}, but it displays only static visualizations at predefined orientations. In contrast, the visualization function in \textbf{fabisearch} produces an elegant network laid over a cortical surface, which can be infinitely adjusted by rotating in 3-dimensions over the x, y, and z axes using the cursor. Accordingly, this provides a level of interaction and exploratory potential that is not possible with simple, flat visualizations. Conveniently, \textbf{fabisearch} can be used with any node coordinate atlas such as the Gordon atlas \citep{Gordon2016} and the Automated Anatomical Labelling atlas \citep{tzourio}, and for any manually uploaded coordinate inputs. Nodes can also be colored based on community membership, which makes it easy to examine complex interactions.  In addition, the scope of the networks can be adjusted by sub-selecting nodes to include in the visualization, which may allow for easier interpretation. Lastly, the visualization function is general; it can also be used on any network data sets without any change points, which increases its applicability.

The rest of this paper is organized as follows. First, we provide a brief overview of the FaBiSearch methodology in Section \ref{sec:models}. In Section \ref{sec:softoverview} we detail the functionality of \textbf{fabisearch}, while in Section \ref{sec:fmri} we implement \textbf{fabisearch} on a resting-state fMRI data set and show its functionality to estimate and visualize networks between change points using a flexible stand-alone function that has the ability to visualize an interactive, 3-dimensional, brain-specific network. Next, we apply \textbf{fabisearch} to a financial time series data set to showcase its generalizability to other data sources in Section \ref{sec:financial}. Finally, we summarize and conclude our work in Section \ref{sec:summary}. The \textbf{fabisearch} package is made freely available on the Comprehensive R Archive Network (CRAN) at \url{https://cran.r-project.org/package=fabisearch}.

\section{Methodology} \label{sec:models}
In this section, we provide an overview of the Factorized Binary Search (FaBiSearch) method for detecting change points in the network (or clustering) structure of multivariate high-dimensional time series data. We also describe how it estimates networks between each pair of detected change points.

\subsection{Model Setup} \label{sec:NMF}
\begin{figure}[h]
\begin{center}
  \includegraphics[width=0.5\linewidth]{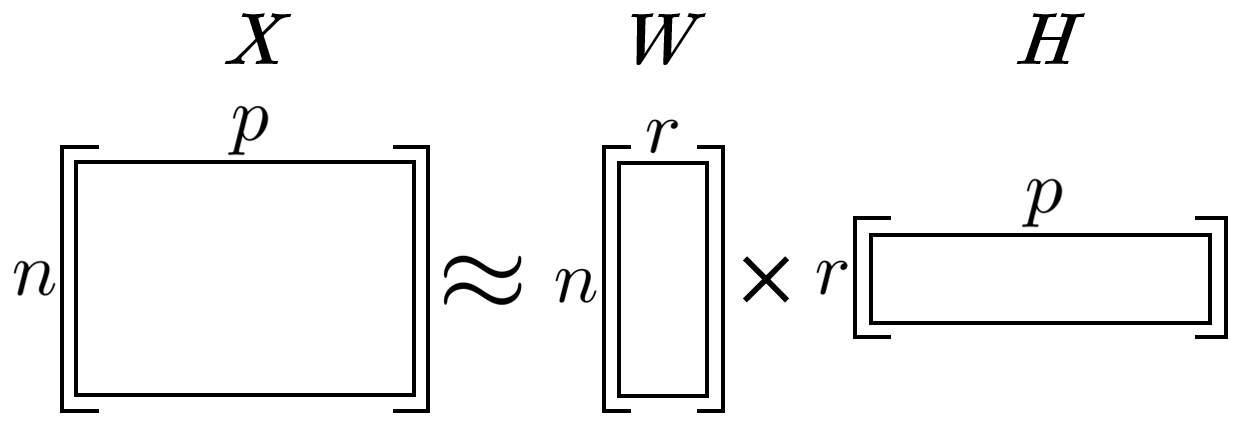}
  \caption{The conceptualization of non-negative matrix factorization (NMF). The input matrix $\boldsymbol{X}$ is approximated by the product of low dimensional factors $\boldsymbol{W}$ and $\boldsymbol{H}$, where $r << min(n,p)$.}
  \label{fig:NMFgeneral}
\end{center}
\end{figure}

Non-negative matrix factorization (NMF) is a matrix factorization technique and an unsupervised dimension reduction method for projecting high-dimensional data sets into lower dimensional spaces \citep{NIPS2000_1861}. Given an $n \times p$ matrix, $\boldsymbol{X}$, where $n$ is the number of independent samples and $p$ is the number of features, NMF approximates $\boldsymbol{X}$, using the product of a $n \times r$ coefficient matrix, $\boldsymbol{W}$, and a $r \times p$ basis matrix, $\boldsymbol{H}$ (Figure \ref{fig:NMFgeneral}). 
The dimensionality of factors $\boldsymbol{W}$ and $\boldsymbol{H}$ is controlled by the factorization rank, $r$, where $r << min(n,p)$. Rank is selected to manage the bias-variance tradeoff, and due to the inherent clustering property of NMF \citep{Li2004,Li2014}, it can be chosen based on the number of unique clusters in $\boldsymbol{X}$.

An important step for comparing the quality of NMF models \citep{NIPS2000_1861} is calculating the goodness of fit between $\boldsymbol{X}$ and the factorization $\boldsymbol{W}\boldsymbol{H}$. We use the generalized Kullback-Leibler divergence (KLD), or \textit{I}-divergence: \citealp{honkela}, based loss,
\begin{equation}
D(\boldsymbol{X}|\boldsymbol{W}\boldsymbol{H}) = \sum_{\substack{ij}}(X_{ij} \log \frac{X_{ij}}{(WH)_{ij}} - X_{ij} + (WH)_{ij}),
\label{eqn:KD}
\end{equation}
\noindent where the asymmetric divergence between the input matrix $\boldsymbol{X}$ and the NMF factorization $\boldsymbol{W} \boldsymbol{H}$ is calculated for each $ij^{th}$ entry.  This and many other measures of loss as well as algorithms for solving NMF are implemented in the R package \textbf{NMF} \citep{Gaujoux2010}. We use the C++ based \fct{nmf} function as the basis for implementing NMF in our \textbf{fabisearch} package. See \fct{?nmf} for more details on methods available in the \textbf{NMF} package, which correspond to the \texttt{algtype} argument in the \textbf{fabisearch} package.  We adjust the NMF definition to a time series context, where the input multivariate (possibly high-dimensional) time series $\boldsymbol{Y} \in \mathbb{R}^{T \times p}_{\geq 0}$ is approximated by the dot product of $\boldsymbol{W}$ and $\boldsymbol{H}$ with $r << min(T, p)$:
\[ 
\boldsymbol{Y}_{T \times p} \approx \boldsymbol{W}_{T \times r} \boldsymbol{H}_{r \times p},
\]
where $T, r,$ and $p$ denote the number of time points, the rank, and the number of time series, respectively. 

The decomposition estimated by NMF is not guaranteed to be unique, hence a similar $Y$ may be generated by different $W$ and $H$ combinations \citep{NIPS2000_1861}. In FaBiSearch, we seek to model the shared linear dependencies across time series using KLD (\ref{eqn:KD}) as a summary statistic of how well $W$ and $H$ model the structure of $Y$. This setup allows us to simultaneously evaluate the conditions in which both factors, $W$ and $H$, materially change in their combined representation of $Y$. In particular, we define a change point as $t^*$, where there is a statistically significant decrease in the KLD when independently fitting the NMF model to time segments before and after $t^*$, or $\{1:t^*\}$ and $\{t^*+1:T\}$, respectively.

\subsection{Initialization and the factorization rank parameter} \label{sec:initrank}
Given the iterative manner in which the factors, $\boldsymbol{W}$ and $\boldsymbol{H}$ are estimated, FaBiSearch is sensitive to the initial values of these factors. To achieve a satisfactory approximation the factorization must be performed over multiple runs, using random initial values for $\boldsymbol{W}$ and $\boldsymbol{H}$. We define the number of runs as $n_{runs}$, which is set to balance accuracy (higher $n_{runs}$) and computational time (lower $n_{runs}$). In the \textbf{fabisearch} package, the argument for the number of runs is denoted as \texttt{nruns} and by default is set to 50, which reasonably balances accuracy and computational time for most applications (see Section~\ref{sec:summary} for more details).

The factorization rank, $r$, is also a key parameter in NMF. For FaBiSearch, it may be specified a priori if known. However, if unknown, we provide a method for calculating the optimal rank that is adapted from \cite{Frigyesi2008} with the only salient deviation being that we choose to permute $\boldsymbol{Y}$ over both rows and columns instead of just over columns, denoted as $\boldsymbol{Y^*}$. Accordingly, we find the global optimal rank, $r_{opt}$, by comparing the change in loss in response to increasing the rank of the original multivariate time series $\boldsymbol{Y}$ to $\boldsymbol{Y^*}$. The first rank where the decrease in loss for $\boldsymbol{Y}$ is less than for $\boldsymbol{Y^*}$ is denoted by $r_{opt}$.

\subsection{Segmentation} \label{sec:binsearch}
FaBiSearch requires a sufficient number of data points to compute the factorization, hence, we define $\delta$, which is a user specified variable, as the minimum distance between candidate change points. Therefore determining a value for this becomes one of statistical stability and robustness of NMF estimates, and this value depends on the context and application. For example, in higher dimensional data and/or noisier regimes, the amount of time points required to obtain a stable estimate may increase. Naturally, this parameter also defines the distance between the beginning (end) of the time series to the first (last) time point to be evaluated. In practice, $\delta$ should be large enough to maintain sufficient stability in the NMF estimates, but also small enough such that change points grouped closely in time are not missed.

\begin{figure}[ht]
\begin{center}
  \includegraphics[width=0.9\linewidth]{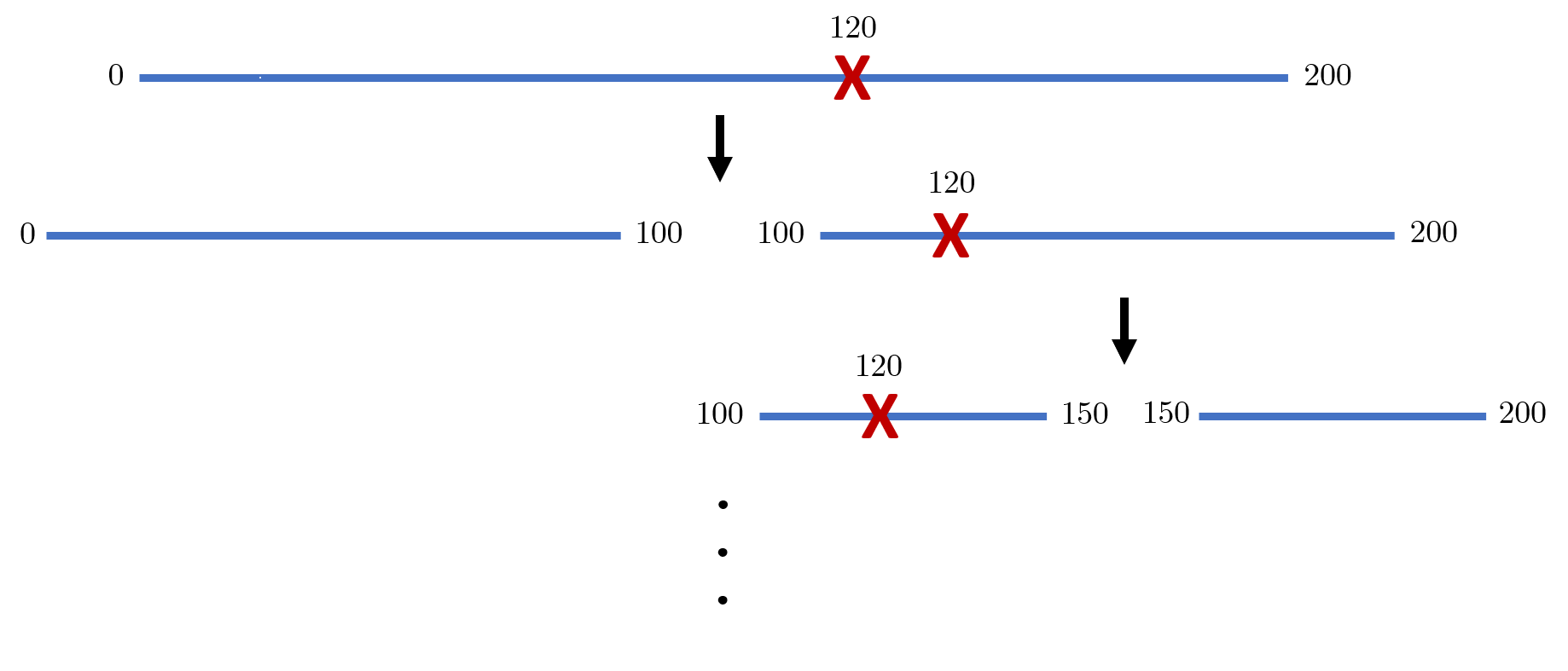}
  \caption{A graphic of the first two iterations of the binary search procedure utilized in FaBiSearch to detect the first candidate change point. The multivariate time series (T = 200) denoted by a blue line, is recursively divided (by approximately half, at each step) to find the true change point ($t^* = 120$) denoted by the red cross.}
  \label{fig:binsearch}
\end{center}
\end{figure}

Binary segmentation is the most common segmentation technique in change point detection, which can be recursively applied to find multiple change points. In this method, all time points from $(\delta)$ to $(T-\delta)$ are sequentially evaluated to find the first candidate change point, which maximizes or minimizes a chosen criterion \citep{Douglas1973ALGORITHMSFT, RAMER1972244, doi:10.1086/620282}. For our applications we are interested in multivariate high-dimensional time series, hence applying NMF to all possible time points $(\delta)$ to $(T-\delta)$ with a reasonable $n_{runs}$ is computationally cumbersome. This behaviour is typical in higher dimension data sets, where each model fitment is expensive, and so binary segmentation is typically too slow to implement. Other methods to speed up change point detection in such a context have been suggested by \cite{kovacs2020optimistic, kovacs2023seeded}. In FaBiSearch, we implement a binary search based method as the optimization technique for efficient segmentation (for more details see \citealp{ondrus2021factorized}). The principle of this technique is based on the intuition that fitments of NMF on data with changing clustering structure result in higher loss, where the segment with higher loss is more likely to contain a change point, which is loosely related to the idea of optimistic search in \cite{kovacs2020optimistic}. To begin, the time series is split into two equal length segments. From here, each segment is separately fit with NMF and the losses of the two segments are calculated. The segment with higher loss is the segment with greater clustering heterogeneity, and therefore greater likelihood of containing a change point.  This process is repeated, and at each step the size of the time segment reduces by approximately half (Figure \ref{fig:binsearch}). This continues until the segment is narrowed down to a single time point, which is then saved as the the first candidate change point, $\hat{q}_{1}$. The time series is then separated into two components, (1:$\hat{q}_{1}$) and ($\hat{q}_{1}$+1:T), and the binary search algorithm is repeated recursively on each component to detect multiple change points.

\subsection{Refitting segments and statistical inference} \label{sec:refit}
FaBiSearch's strategy is to first overestimate the number of change points by recursively finding candidate change points, and then pruning the results through a refitting and statistical inference procedure. To determine which change points to keep, we consider a permutation test \citep{fisher1936design, collingridge2013primer} for each candidate. To begin, we define a set $\hat{Q}$, which includes all candidate change points $\hat{q}_{1}, ...,\hat{q}_{k}$ (detected as depicted in Section \ref{sec:binsearch}) that have been arranged in ascending order. Then, we define a set, $W = \{1, \hat{Q}, T\}$, from $\hat{Q}$, and the beginning and the end of the time series ($1$ and $T$, respectively). For each candidate change point, $\hat{q}_{i}$, we define the left and right boundaries
\[
b^{L}_{i} = (w_{i}:w_{i+1});~~~~b^{R}_{i} = (w_{i+1}+1:w_{i+2}) \text{.}
\]
\noindent Next, we define the data segment for $\hat{q}_{i}$ as $\boldsymbol{Z}_{i} \gets \{\boldsymbol{Y}_{t}\}_{b^{L}_{i}\cup b^{R}_{i}}$ and corresponding sub-segments $\boldsymbol{Z}_{i}^{L}$ and $\boldsymbol{Z}_{i}^{R}$ which capture data before and after the candidate change point $\hat{q}_{i}$ as
\[
\boldsymbol{Z}_{i}^{L} \gets \{\boldsymbol{Y}_{t}\}_{b^{L}_{i}} \text{ ; } \boldsymbol{Z}_{i}^{R} \gets \{\boldsymbol{Y}_{t}\}_{b^{R}_{i}} \text{.}
\]

\noindent After this, we re-fit NMF to data sub-segments $\boldsymbol{Z}_{i}^{L}$ and $\boldsymbol{Z}_{i}^{R}$. The loss (difference between the original data and its reconstruction) of these two sub-segments is summed for each fitment over $n_{reps}$, generating a distribution of losses, $l_{i} = \{l_{i,1},...,l_{i,n_{reps}}\}$. To then determine whether to keep or remove the candidate change point, we compare this distribution to a reference distribution. This reference distribution is generated by first taking the partition, $\boldsymbol{Z}_{i}$, and permuting across time points to disrupt the temporal organization of the data structure. For each permutation in $n_{reps}$, we re-fit NMF and sum the loss of the two sub-segments, generating reference distribution $l_{i}^{*} = \{l_{i,1}^{*},...,l_{i,n_{reps}}^{*}\}$. By default, statistical inference between these two distributions is performed using a one-sided, two sample $t$-test assuming unequal variances (Welch's $t$-test). Recall previously the intuition that time series with a change point have inconsistent clustering structure and thus exhibit higher loss when fitting NMF. Consequently, we are interested in the following hypothesis:
\[
\begin{array}{c}
H_{0} : \mu(l_{i}) \geq \mu(l^{*}_{i}) \\
H_{a} : \mu(l_{i}) < \mu(l^{*}_{i}).
\end{array}
\]

In the \textbf{fabisearch} package, we also provide two non-parametric tests, namely the Wilcoxon signed-rank test and the Kolmogorov-Smirnov test, to compare the re-fitted distribution, $l_{i}$, to the reference distribution, $l^{*}_{i}$. The null hypothesis for Wilcoxon signed-rank test is similar to the two sample $t$-test except the test is on the medians of the distribution, and the hypothesis for the Kolmogorov-Smirnov test is as follows:
\[
\begin{array}{c}
H_{0} : l_{i} \text{ and } l^{*}_{i} \text{ are identically distributed} \\
H_{a} : \text{The cumulative distribution function of } l_{i} \text{ lies above } l^{*}_{i}  \text{.}
\end{array}
\]

\noindent Since FaBiSearch may find multiple change points in the input time series $\boldsymbol{Y}$, we adjust the $p$-value for each statistical test for multiple comparisons using the \cite{benjamini} method. For the statistical tests, if we reject $H_{0}$, we retain the change point $\hat{q}_{i}$, otherwise we remove it. 

\subsection{Estimating stationary networks} \label{sec:networkestimation}
After the change points have been detected, stable networks between the change points can be estimated for visualization and interpretation purposes, using various methods such as correlation or precision matrices.  Here, however, we introduce an NMF-based method for computing an adjacency matrix for data between change points.  The procedure proceeds as follows: the first step is to re-apply NMF to each stationary block of data. Then, using values in the coefficient matrix, $\boldsymbol{H}$, the cluster membership of each time series (or variable) is determined.  Specifically, each column in $\boldsymbol{H}$ is assigned to cluster $r$ which has the highest coefficient value. Then, for each run in $n_{runs}$, an adjacency matrix,
\begin{equation*}
\boldsymbol{A}_{ij} = \begin{cases}
1, & \mbox{if } i, j \mbox{ are in the same cluster};\\
0, & \mbox{otherwise}
 \end{cases}
\end{equation*}
\noindent is generated denoting cluster membership of each component in the multivariate time series, $\boldsymbol{Y}$. This information is combined into a consensus matrix, which is computed by averaging the individual adjacency matrices:
\[
\boldsymbol{C}_{ij} = \mu(\boldsymbol{A}_{ij}^1 \text{ ,..., } \boldsymbol{A}_{ij}^{n_{run}}), ~~ 0 \leq \boldsymbol{C}_{ij} \leq 1 \text{.}
\]

\noindent The main advantage of this procedure is that it combines results across $n_{runs}$, which may separately have high variance, into a matrix where each entry denotes the probability of two nodes being in the same cluster.  This procedure is similar to stability selection in \cite{meinshausen}, and the bootstrapping in \cite{zhu2018}.  Hence, there are two possible methods for defining the final adjacency matrix from this consensus matrix. The first method entails applying hierarchical clustering to classify cluster membership amongst nodes from the consensus matrix, and cutting the resulting tree at a predetermined number of clusters. The second method entails defining an adjacency matrix from the consensus matrix using a prespecified threshold, $\lambda$, which controls the sparsity of the final adjacency matrix, that is,
\begin{equation*}
\boldsymbol{A}_{ij} = \begin{cases}
1, & \mbox{if } \boldsymbol{C}_{ij} > \lambda;\\
0, & \mbox{otherwise}.
 \end{cases}
\end{equation*}

%% -- Software Overview -------------------------------------------
\section{Software Overview} \label{sec:softoverview}

In this section, we provide an overview of the functions included in \textbf{fabisearch}, how they connect to the methodology, and the related arguments for the functions.

\subsection{Shared arguments} \label{sec:shared}

We first delineate four arguments which have the same definitions and uses in the functions \fct{opt.rank}, \fct{detect.cps}, and \fct{est.net}.

\begin{itemize}
    \item \texttt{Y} is the input multivariate time series in matrix format, with variables organized in columns and time points in rows. All entries in \texttt{Y} must be positive.
    \item \texttt{nruns} is the number of runs, or random restarts, when computing NMF. This value should scale appropriately to the size of the input matrix \texttt{Y}; larger data sets require more random starts to obtain convergence to a stable estimation of the NMF parameters.
    \item \texttt{rank} is a hyperparameter for NMF and is used to balance the bias-variance tradeoff when estimating the model parameters. \texttt{rank} can be set to a positive integer value. By default, \texttt{rank} is set to \texttt{NULL}, and the procedure \fct{opt.rank} as described in Section \ref{sec:initrank} is used to find the optimal rank for \texttt{Y}.
    \item \texttt{algtype} denotes the NMF algorithm. The methods available correspond to those available in \cite{Gaujoux2010}'s \textbf{NMF} package. By default it is set to the \texttt{"brunet"} algorithm, see \fct{?nmf} for other available methods.
\end{itemize}

\subsection{Finding optimal rank}

We provide an automated method for finding the optimal rank of the multivariate time series, $\boldsymbol{Y}$, through the

\begin{center}
\texttt{opt.rank(Y, nruns = 50, algtype = "brunet")}
\end{center}

\noindent function. All arguments are described in Section \ref{sec:shared}. \fct{opt.rank} returns a numeric integer denoting the optimal rank.

\subsection{Change point detection}

Change point detection using the FaBiSearch method is carried out by the 

\begin{center}
\texttt{detect.cps(Y, mindist = 35, nruns = 50, nreps = 100, alpha = NULL, }

\texttt{rank = NULL, algtype = "brunet", testtype = "t-test")}
\end{center}

\noindent function. Arguments \texttt{Y}, \texttt{nruns}, \texttt{rank}, and \texttt{algtype} are described in Section \ref{sec:shared}, and the remaining arguments are characterized as follows:

\begin{itemize}
    \item \texttt{mindist} is the minimum distance between change points, and also corresponds to the minimum number of time points to be included in the NMF estimation. By default it is set to 35. It should be large enough such that adequate precision is attained when estimating parameters, but small enough to account for possibly multiple change points being grouped closely in time (see Section \ref{sec:binsearch}).
    \item \texttt{nreps} corresponds to the number of repetitions for the statistical inference step. It dictates the number of permutations generated for the statistical inference test on the candidate change points (Welch's $t$-test, the Wilcoxon signed-rank test, or the Kolmogorov-Smirnov test) and should be large enough such that adequate statistical power is attained.
    \item \texttt{alpha} is the significance level. It can be set to a positive integer denoting the $\alpha$ value to use in the Welch's $t$-test, the Wilcoxon signed-rank test, or the Kolmogorov-Smirnov test. By default, it is set to \texttt{NULL}, which returns the \textit{p}-value of the test.
    \item \texttt{testtype = "t-test"} is a character string, which defines the type of statistical test to use during the inference procedure. By default it is set to ``t-test''. The other options are ``ks'' and ``wilcox'' which correspond to the Kolmogorov-Smirnov and the Wilcoxon signed-rank tests, respectively.
\end{itemize}

\fct{detect.cps} returns a list with the following three components:

\begin{itemize}
    \item \texttt{\$rank} is the rank used for NMF estimation.
    \item \texttt{\$change\_points} are the results of the procedure summarized in a table. Each row in the matrix corresponds to a candidate change point, where columns \texttt{T} and \texttt{stat\_test} correspond to the time of the change point and the result of the statistical inference test, respectively. If \texttt{alpha} is a positive real number, this value is used as the significance level for the statistical inference test, and a boolean value is returned based on its statistical significance. Conversely, if the \texttt{alpha} argument is set to the string \texttt{"p-value"}, then the \textit{p}-value of the statistical test is returned.
    \item \texttt{\$compute\_time} is the computational time of the procedure, saved as a \texttt{difftime} object.
\end{itemize}

\subsection{Estimating stationary networks} \label{sec:est.netFUNCTION}
As described in Section~\ref{sec:networkestimation}, methods for estimating stationary networks between change points are implemented in \textbf{fabisearch} using the 

\begin{center}
\texttt{est.net(Y, lambda, nruns = 50, rank = "optimal", algtype = "brunet", changepoints = NULL)}
\end{center}

\noindent function. Arguments \texttt{Y}, \texttt{nruns}, \texttt{rank}, and \texttt{algtype} are described in Section \ref{sec:shared}, and the remaining arguments are characterized as follows:

\begin{itemize}
    \item \texttt{lambda} has two purposes: first, it is used to specify the method for calculating an adjacency matrix (clustering or thresholding). Second, it is used to denote a value in the selected process (that is, the number of clusters or the cutoff value, respectively). If \texttt{lambda} is a positive integer, then the clustering based method is used and the integer corresponds to the number of clusters. Conversely, if \texttt{lambda} is a positive real number $<1$, then the thresholding method is performed and the \texttt{lambda} value is used as the cutoff for the consensus matrix. \texttt{lambda} may also be a vector of either cluster or threshold values, in which case the output is a list of results where each component corresponds to a \texttt{lambda} value in the vector.
    \item \texttt{changepoints} is a vector of positive integers with default value equal to NULL. It is used to specify whether change points exist in the input multivariate time series, \texttt{Y}, and thus whether \texttt{Y} should be split into multiple stationary segments and networks estimated separately for each. If \texttt{changepoints}, say c(100, 200), are specified, \texttt{Y} is split at time points $100$ and $200$, corresponding to 3 stationary segments. For each stationary segment an adjacency matrix is estimated sequentially, and returned as a list where each component corresponds to a stationary segment.
\end{itemize}

\fct{est.net} returns a square $p \times p$ adjacency matrix in the following format:

\begin{equation*}
\boldsymbol{A} = \begin{cases}
1, & \mbox{if two nodes share an edge};\\
0, & \mbox{otherwise}.
 \end{cases}
\end{equation*}

\subsection{Brain network visualization} \label{sec:visual}

To visualize the stationary networks described in Section \ref{sec:networkestimation}, we create an interactive 3-dimensional plot for brain-specific networks through the 

\begin{center}
\texttt{net.3dplot(A, ROIs = NULL, colors = NULL, coordROIs = NULL)}
\end{center}

\noindent function. The input data, \texttt{A}, is a $p \times p$ adjacency matrix stored as a numerical matrix in the following format:

\begin{equation*}
\boldsymbol{A} = \begin{cases}
1, & \mbox{if two nodes share an edge};\\
0, & \mbox{otherwise}.
 \end{cases}
\end{equation*}

\noindent \texttt{A} in \fct{net.3dplot} can be any adjacency matrix and does not have to be the result of the change point method discussed in the previous sections, hence \fct{net.3dplot} is flexible and extends beyond the included change point estimation method. The remaining input arguments are described as follows:

\begin{itemize}
    \item \texttt{ROIs} is either a vector of character strings specifying the communities to plot, or a vector of integers specifying which ROIs to plot by their ID. By default it is set to \texttt{NULL} and all communities and ROIs are plotted. Communities available for the Gordon atlas include: ``Default'', ``SMhand'', ``SMmouth'', ``Visual'', ``FrontoParietal'', ``Auditory'', ``None'', ``CinguloParietal'', ``RetrosplenialTemporal'', ``CinguloOperc'', ``VentralAttn'', ``Salience'', and ``DorsalAttn''.
    \item \texttt{colors} sets the color of nodes, in hex code format, in the final plot based on community membership. If a vector of \texttt{communities} is specified, then the $n^{th}$ community is assigned the $n^{th}$ color in the \texttt{colors} vector.
    \item \texttt{coordROIs} is a dataframe of community tags and Montreal Neurological Institute (MNI) coordinates for regions of interest (ROIs) to plot, which is by default set to \texttt{NULL} and uses the Gordon atlas \citep{Gordon2016}. See \texttt{?gordon.atlas} for an example using the Gordon atlas. The format of the dataframe is as follows:
    \begin{center}
    \begin{tabular}{|c|c|c|c|}
        \hline
        Communities & x.mni & y.mni & z.mni \\
        \hline
        $<$string$>$ & $<$double$>$ & $<$double$>$ & $<$double$>$ \\
        \vdots & \vdots & \vdots & \vdots \\
        \hline
    \end{tabular}
    \end{center}
    The first column is a string of community labels, and the subsequent three columns are the x, y, and z coordinates in MNI space, respectively. If communities are not applicable to the atlas, the first column is defined by \texttt{NA} (e.g., in the \texttt{AALatlas}). Using this format, any atlas/node coordinate system can be used with the \fct{net.3dplot} function. 
\end{itemize}

The dimension of \texttt{A} must be congruent with the node atlas defined by \texttt{coordROIs}. For example, the default \texttt{gordon.atlas} has $p = 333$, and so \texttt{A} must be a $333 \times 333$ matrix. \fct{net.3dplot} opens and displays brain networks in an interactive 3-dimensional RGL window, which can be interacted with by clicking and dragging the cursor.

%% -- Illustrations ------------------------------------------------------------

%% - Virtually all JSS manuscripts list source code along with the generated
%%   output. The style files provide dedicated environments for this.
%% - In R, the environments {Sinput} and {Soutput} - as produced by Sweave() or
%%   or knitr using the render_sweave() hook - are used (without the need to
%%   load Sweave.sty).
%% - Equivalently, {CodeInput} and {CodeOutput} can be used.
%% - The code input should use "the usual" command prompt in the respective
%%   software system.
%% - For R code, the prompt "R> " should be used with "+  " as the
%%   continuation prompt.
%% - Comments within the code chunks should be avoided - these should be made
%%   within the regular LaTeX text.

%================================================================================================
\section{Resting-state fMRI Example} \label{sec:fmri}
This resting-state fMRI data set includes 25 participants scanned at New York University over three visits (\url{http://www.nitrc.org/projects/nyu_trt}).  For each visit, participants were asked to relax, remain still, and keep their eyes open.  A Siemens Allegra 3.0-Tesla scanner was used to obtain the resting-state scans for each participant.  Each visit consisted of 197 contiguous EPI functional volume scans with time repetition (TR) of 2000ms, time echo (TE) of 25ms, flip angle (FA) of 90$^{\circ}$, 39 number of slices, matrix of $64\times64$, field of view (FOV) of 192mm, and voxel size of $3\times3\times3$mm$^3$. Software packages \texttt{AFNI }(\url{http://afni.nimh.nih.gov/afni}) and \texttt{FSL }(\url{http://www.fmrib.ox.ac.uk}) were used for preprocessing. Motion was corrected using \texttt{FSL}'s \texttt{mcflirt} (rigid body transform, cost function normalized correlation, and reference volume the middle volume). Normalization into the Montreal Neurological Institute (MNI) space was performed using \texttt{FSL}'s \texttt{flirt} (affine transform, cost function, mutual information). Probabilistic segmentation was conducted to determine white matter and cerebrospinal fluid (CSF) probabilistic maps and was obtained using \texttt{FSL}'s \texttt{fast} with a threshold of 0.99. Nuisance signals (the six motion parameters, white matter signals, CSF signals, and global signals) were removed using \texttt{AFNI}'s \texttt{3dDetrend}. Volumes were spatially smoothed using a Gaussian kernel and FWHM of 6mm with \texttt{FSL}'s \texttt{fslmaths}. We used the work of \cite{Gordon2016} to determine the ROI atlas, which results in the cortical surface being parcellated into 333 areas of homogenous connectivity patterns, and the time course for each is determined by averaging the voxels within each region for each subject. Regional time courses were then detrended and standardized to unit variance. Lastly, a fourth-order Buttterworth filter with a 0.01-0.10 Hertz pass band was applied.  Hence, in summary, we have a high-dimensional time series data set with $T=197$ and $p=333$. For more details on the data set, see \cite{xu}.

\subsection{Data summary and change point detection} \label{sec:datacpd}
We focus on the second scan of the first subject of the resting-state fMRI data set for display purposes. To begin, we  inspect the data by calculating summary statistics and plotting individual time series for the first 4 ROIs (or nodes):

\begin{verbatim}
R> data("gordfmri", package = "fabisearch")
R> print(gordfmri[1:10,1:4])
           1         2         3         4
1  101.38913  99.42839 100.52694 102.54848
2   97.81678  98.92552  99.19077 101.17978
3   98.19218  99.78406  97.61275 101.00244
4  102.66680  99.15784  98.88478 102.29753
5  102.12487 100.05531  97.08151  99.14587
6  104.41834  99.33773  97.17439  98.52235
7  101.86739  99.48253  96.97930 100.00666
8   98.79015  99.78261  99.77578  95.95149
9  100.03848 100.08721  98.35715 101.50598
10  99.73903  97.77725  98.39061  99.86871

R> summary(gordfmri[,1:4])
       1                2                3                4         
 Min.   : 94.42   Min.   : 96.98   Min.   : 96.98   Min.   : 95.54  
 1st Qu.: 98.60   1st Qu.: 99.16   1st Qu.: 99.03   1st Qu.: 98.78  
 Median : 99.74   Median : 99.90   Median :100.06   Median : 99.81  
 Mean   :100.00   Mean   :100.00   Mean   :100.00   Mean   :100.00  
 3rd Qu.:101.39   3rd Qu.:100.96   3rd Qu.:100.80   3rd Qu.:101.31  
 Max.   :106.73   Max.   :103.64   Max.   :103.88   Max.   :104.67
 
R> par(mfrow=c(4,1))
R> for(i in 1:4){
+  plot(gordfmri[,i], type = "l", cex.lab = 1.5, cex.axis = 1.5,
+    xlab = "Time", ylab = paste("Node", i))
+}
\end{verbatim}

\begin{figure}[ht]
\begin{center}
  \includegraphics[width=1\linewidth]{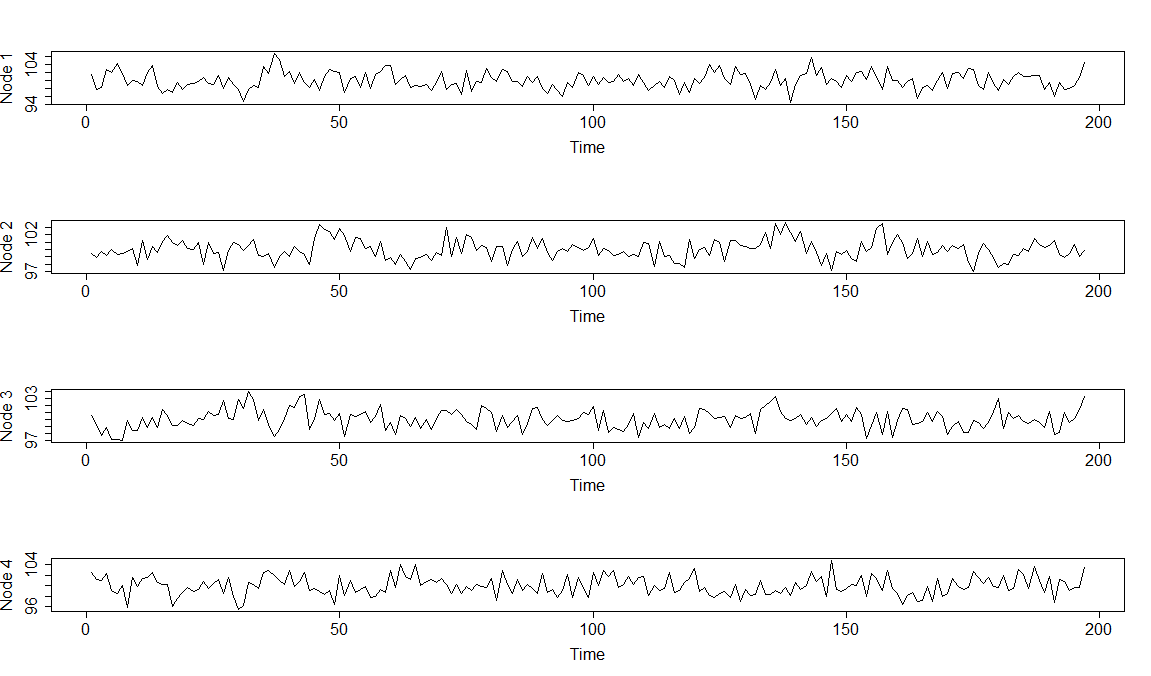}
  \caption{Plots of the first four ROI (or column, or node) time series of the \texttt{gordfmri} data. \texttt{gordfmri} is processed from the second scan of the first subject from the NYU test-restest resting-state fMRI data set.}
  \label{fig:fmrifirst4}
\end{center}
\end{figure}

Since we are interested in finding change points in this data set, we must consider the input arguments for \fct{detect.cps}. We assume that the number of clusters is unknown a priori, hence, we keep \texttt{rank} at its default value, \texttt{optimal}. Additionally, we also keep \texttt{alpha} at its default value, \texttt{NULL}, which returns the $p$-values for each change point. We use default values for all other arguments.

\begin{verbatim}
R> set.seed(12345)
R> fmrioutput = detect.cps(gordfmri)
\end{verbatim}

As \fct{detect.cps} runs, it prints updates in the console. We show these console messages below:

\begin{verbatim}
[1] "Finding optimal rank"
[1] "Optimal rank: 7"
[1] "35 : 162"
[1] "35 : 99"
[1] "35 : 67"
[1] "35 : 51"
[1] "35 : 43"
[1] "35 : 39"
[1] "35 : 37"
[1] "35 : 36"
[1] "Change Point At: 35 , Delta Loss: -26.3978932820244"
[1] "70 : 162"
[1] "70 : 116"
[1] "70 : 93"
[1] "70 : 82"
[1] "70 : 76"
[1] "70 : 73"
[1] "70 : 72"
[1] "70 : 71"
[1] "Change Point At: 70 , Delta Loss: -14.7458419916325"
[1] "105 : 162"
[1] "133 : 162"
[1] "133 : 148"
[1] "133 : 141"
[1] "136 : 141"
[1] "136 : 139"
[1] "136 : 138"
[1] "136 : 137"
[1] "Change Point At: 136 , Delta Loss: -18.0354425518596"
[1] "Refitting split at 35"
[1] "Refitting split at 70"
[1] "Refitting split at 136"
[1] "Permuting split at 35"
[1] "Permuting split at 70"
[1] "Permuting split at 136"
\end{verbatim}

Since we specified that the factorization parameter, \texttt{rank}, to be optimally determined, the first set of messages relate to this process. The procedure finds an optimal \texttt{rank} equal to 7, and then the function proceeds to apply the FaBiSearch method to the multivariate high-dimensional time series. Each subsequent message corresponds to the time indices being evaluated at that moment, where, as described in Section \ref{sec:binsearch}, time indices are iteratively halved to find candidate change points. Consequently, the time indices being evaluated reduce with each iteration. Once a candidate change point has been detected (e.g., time point 35), this process is applied recursively to the child matrices whose boundaries are defined by this new candidate change point (e.g., time points 70:162). As soon as this search has been exhausted and all candidate change points have been detected (e.g., time points 35, 70, 136), the refitting, permutation, and statistical inference procedures are performed (Section \ref{sec:refit}). \fct{detect.cps} saves the output as a list with 3 components:

\begin{verbatim}
R> fmrioutput
$rank
[1] 7

$change_points
    T    stat_test
1  35 8.956153e-07
2  70 3.987376e-07
3 136 5.344967e-02

$compute_time
Time difference of 12.79541 mins
\end{verbatim}

In the second element of the output list above, each row contains a candidate change point and its associated $p$-value. Using a stringent significance level of $\alpha = 0.001$, we can determine which of the candidates we retain as change points.

\begin{verbatim}
R> finalcpt = fmrioutput$change_points[fmrioutput$change_points[,2] 
+   < 0.001, 1]
R> finalcpt
[1]  35 70
\end{verbatim}

The change points detected by FaBiSearch occur at time points 35 and 70, which correspond to changes in the network clustering structure of the particular subject. Even at rest, fMRI time series data shows evidence of non-stationarity (\citealp{delamillieure, cribben2017,anastasiou} to name just a few) as subjects drift between functional modes or states of thought and attentiveness which explains the changing network dynamics over the experiment.

\subsection{Estimating stationary networks} \label{sec:eststatnet}

\begin{figure}[h]
\begin{center}
  \includegraphics[width=1\linewidth]{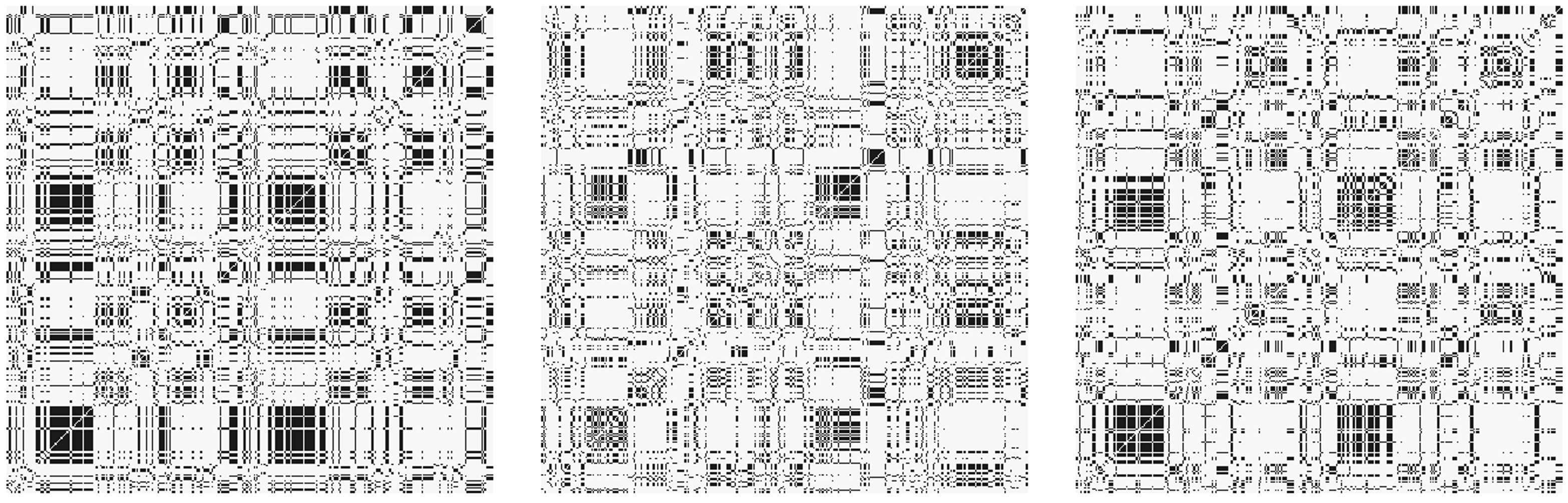}
  \caption{The adjacency matrices for the three stationary networks in the \texttt{gordfmri} data set using the clustering based network estimation method. From left to right the adjacency matrices correspond to stationary networks for time points 1:35, 36:69, and 70:197, respectively. For each adjacency matrix, nodes are numbered 1:333 advancing from left to right in columns, and from bottom to top in rows.   The \texttt{gordfmri} data is processed from the second scan of the first subject from the NYU test-restest resting-state fMRI data set.}
  \label{fig:clustnetheatmaps}
\end{center}
\end{figure}

For the second scan of the first subject of the resting-state fMRI data set in the previous section, we detected change points at time points $t = 35, 70$. With two change points, we have three stationary networks to estimate. We specify these two change points as the \texttt{changepoints} argument in \fct{est.net} so that the data, \texttt{gordfmri}, is segmented appropriately.  Given we identified an optimal rank of 7 for \texttt{gordfmri} in Section \ref{sec:datacpd}, we use this as the number of clusters in each of the stationary blocks. Consequently, we set \texttt{lambda = 7} and \texttt{rank = 7} in \fct{est.net}, and save the results in \texttt{clust.net}. We carry over and use \texttt{finalcpt} to specify the change points for the argument \texttt{changepoints}. From the two methods for estimating networks in the stationary blocks (Section \ref{sec:networkestimation}), we first explore the clustering based method.

\begin{verbatim}
R> clust.net = est.net(gordfmri, lambda = rankest, rank = rankest,
+                     changepoints = finalcpt)
\end{verbatim}

The output from this is a list with three components, where each component is an adjacency matrix for each corresponding stationary segment. The components themselves are organized sequentially based on time indices (e.g., components 1, 2, and 3 correspond to time points 1:35, 36:70, and 71:197, respectively).  We visualize the resulting networks in Figure \ref{fig:clustnetheatmaps} as adjacency matrices using the \fct{heatmap} function, where \texttt{n} is the $n^{th}$ stationary network in the \texttt{clust.net} list:
\begin{verbatim}
R> heatmap(clust.net[[n]], col = grey(c(0.97,0.1)), symm = TRUE,
+         Colv = NA, Rowv = NA, labRow = NA, labCol = NA)
\end{verbatim}

These adjacency matrices show a strong block diagonal structure, since the clustering based method naturally organizes clusters into separate blocks in each adjacency matrix.  Additionally, we can estimate networks from each of these stationary blocks using the thresholding method (Section \ref{sec:est.netFUNCTION}). Again, we use the \fct{est.net} function and set \texttt{rank = 7}, however, we need to determine an appropriate cutoff value, \texttt{lambda}. Typically, this is unknown a priori and varies based on the estimated consensus matrix. Given the large number of nodes ($p = 333$), there are 55278 possible edges (or upper triangular elements in the adjacency matrix). We seek to find a cutoff which incorporates an adequate amount of variability (small \texttt{lambda}), but minimizes the addition of noise and maximizes sparsity for easier interpretability (large \texttt{lambda}).  One approach is determine an appropriate cutoff value (or the number of edges in the graph), \texttt{lambda}, is to use a heuristic, such as the elbow method, to find the value where increasing \texttt{lambda} has diminishing marginal returns for improving sparsity. To do so, we first define \texttt{lambda} as a vector of possible cutoff values, which in this case is a sequence from 0.01 to 0.99 with 0.01 sized steps. Results are saved as a list in \texttt{thresh.net}.

\begin{verbatim}
R> lambda = seq(0.01, 0.99, 0.01)
R> thresh.net = est.net(gordfmri, lambda = lambda, rank = rankest,
+                      changepoints = finalcpt)
\end{verbatim}

This list however, has two levels. The first level has three elements, where each element corresponds to the stationary network evaluated, just as before. The second level has 99 elements, where each element corresponds to the adjacency matrix at a specific value of \texttt{lambda} (e.g., 0.01, 0.02, 0.03,..., 0.99). We then find an appropriate cutoff value by examining a plot of the \texttt{lambda} value against the number of edges (or elements in the upper triangular part of the adjacency matrix). We save the number of edges in the vector \texttt{edges} as we loop through different values of \texttt{lambda}. For each $n^{th}$ stationary network, we plot the results.

\begin{verbatim}
R> par(mfrow = (c(1,3)))
R> for(n in 1:3){
+   edges = c()
+   for(i in 1:length(lambda)){
+     curr.net = thresh.net[[n]][[i]]
+     edges = c(edges, sum(curr.net[upper.tri(curr.net)]))
+   }
+   plot(lambda, edges, cex.lab = 2, cex.axis = 2)
+ }
R> dev.off()
\end{verbatim}

\begin{figure}[h]
\begin{center}
  \includegraphics[width=1\linewidth]{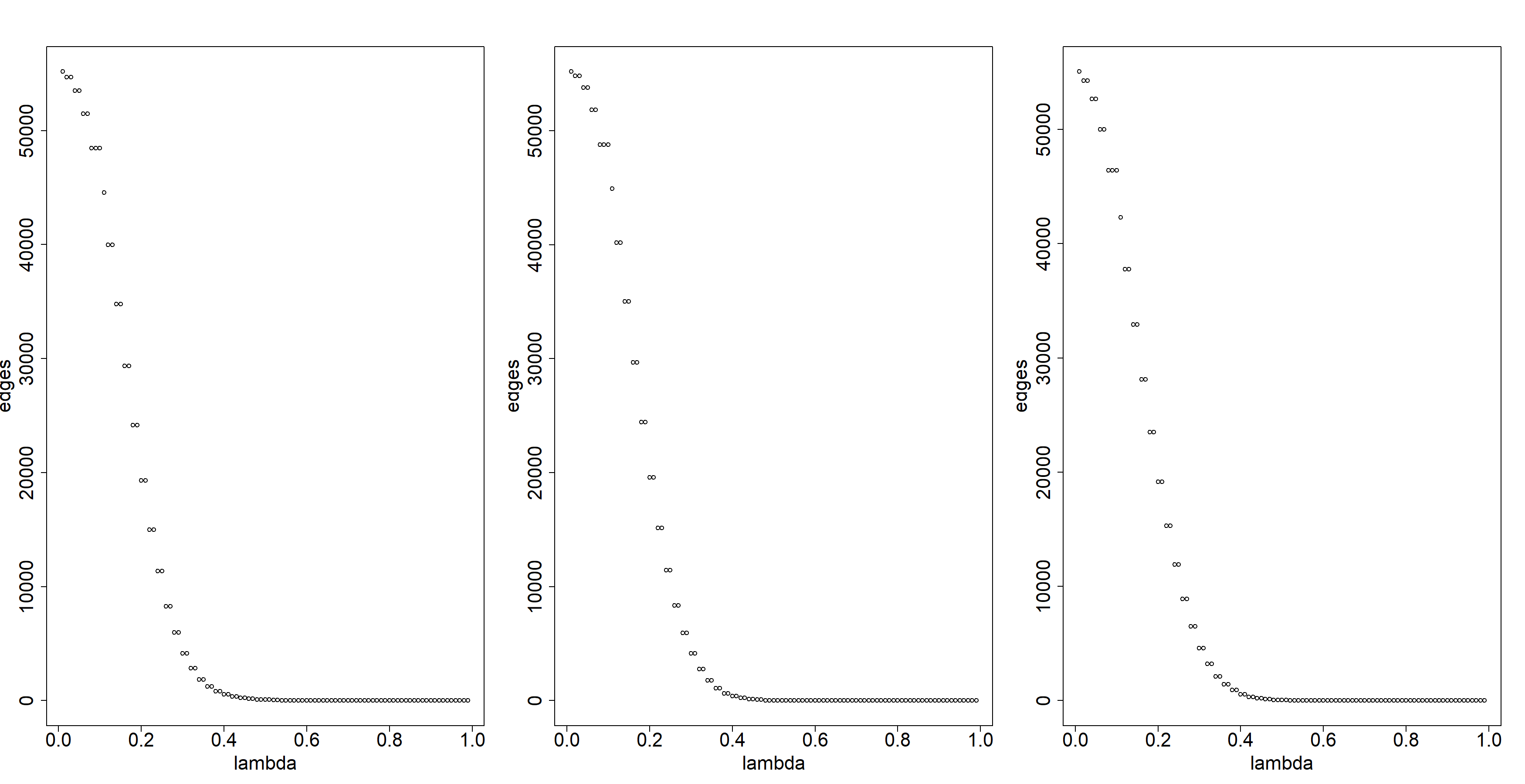}
  \caption{Plots of \texttt{lambda} values against the number of edges  (or elements in the upper triangular part of the adjacency matrices) across stationary networks in the \texttt{gordfmri} data set. From left to right, the plots correspond to the networks for time points 1:35, 36:70, and 71:197, respectively.   The \texttt{gordfmri} data is processed from the second scan of the first subject from the NYU test-restest resting-state fMRI data set.}
  \label{fig:threshnetedges}
\end{center}
\end{figure}

\begin{figure}[h]
\begin{center}
  \includegraphics[width=1\linewidth]{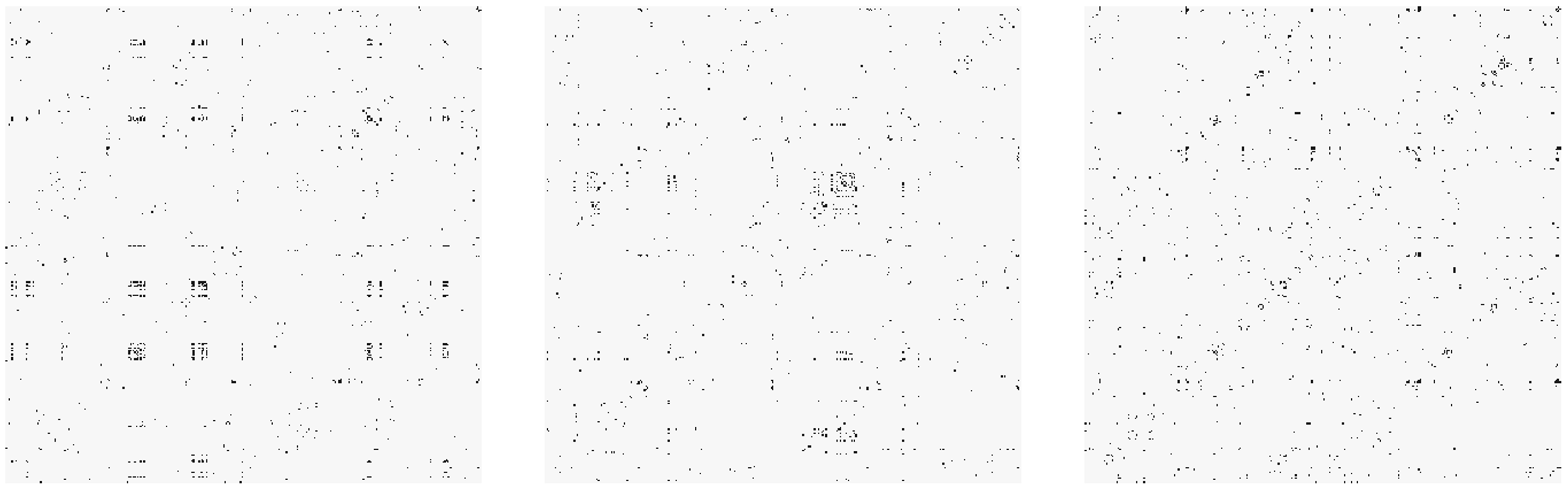}
  \caption{The adjacency matrices for the three stationary networks in the \texttt{gordfmri} data set using the threshold based network estimation method. From left to right the adjacency matrices correspond to stationary networks for time points 1:35, 36:70, and 71:197, respectively. For each adjacency matrix, nodes are numbered 1:333 advancing from left to right in columns, and from bottom to top in rows.  The \texttt{gordfmri} is processed from the second scan of the first subject from the NYU test-retest resting-state fMRI data set. }
  \label{fig:threshnetheatmaps}
\end{center}
\end{figure}

As depicted in Figure \ref{fig:threshnetedges}, an optimal cutoff point is at approximately \texttt{lambda = 0.4} (element 40 in the second level of the \texttt{curr.net} list), hence we use this for computing the new adjacency matrix. For the $40^{th}$ element in the \texttt{curr.net} list, we visualize the adjacency matrices for the corresponding network (Figure \ref{fig:threshnetheatmaps}). Compared to the clustering method, the thresholding method provides sparser adjacency matrices. This highlights the main benefit of this approach compared to the clustering method. It allows for sparsity to be controlled using \texttt{lambda}. Some block diagonal organization remains, however, compared to the clustering method  (Figure \ref{fig:clustnetheatmaps}), it is less prominent.

\begin{verbatim}
R> heatmap(thresh.net[[n]][[40]], col = grey(c(0.97,0.1)), symm = TRUE,
+         Colv = NA, Rowv = NA, labRow = NA, labCol = NA)
\end{verbatim}

\subsection{3-dimensional brain network visualization}
\begin{figure}[h]
\begin{center}
  \includegraphics[width=1\linewidth]{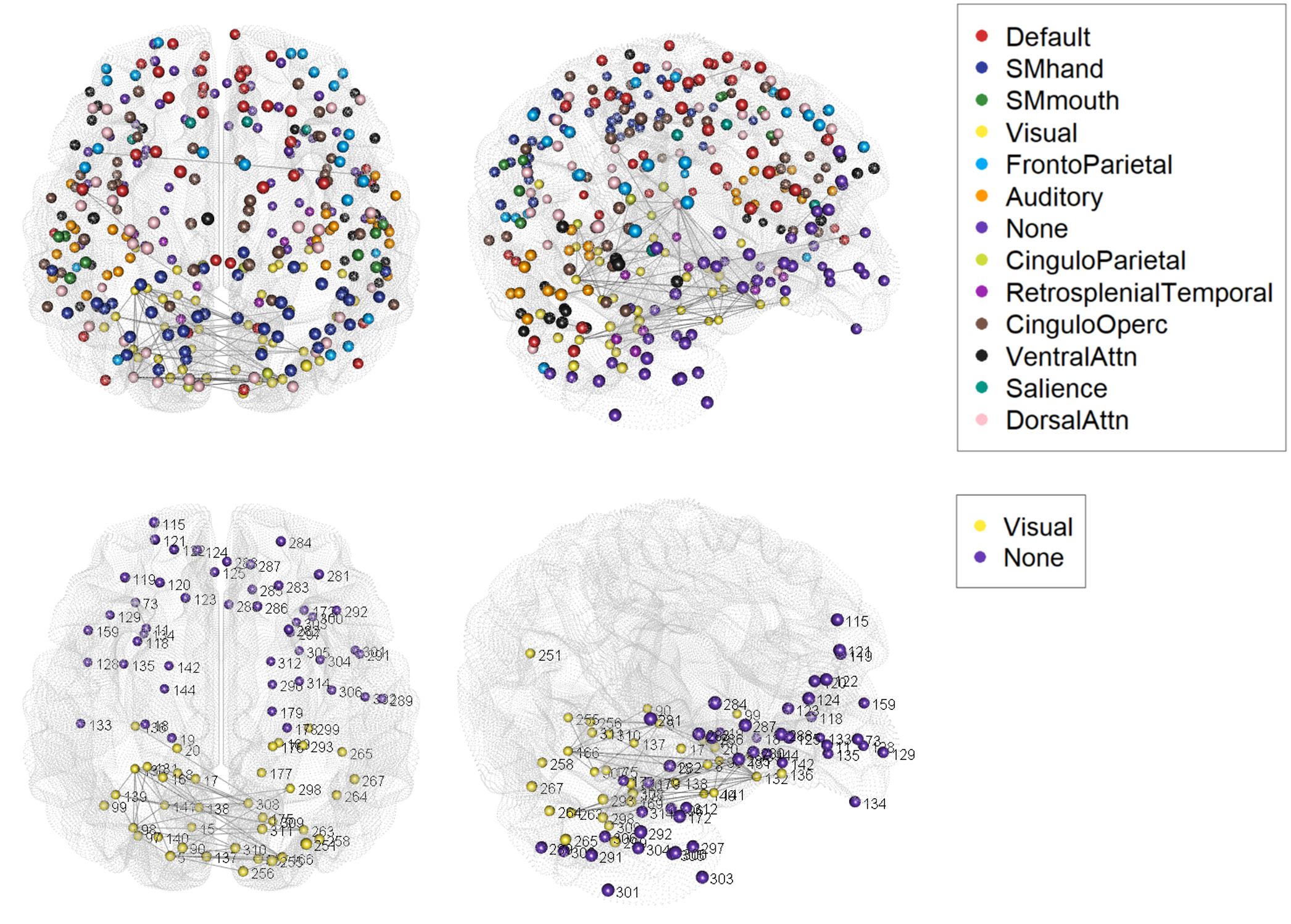}
  \caption{Visualizations of the brain networks of the first stationary block of the \texttt{gordfmri} data, using the \fct{net.3dplot} function. Nodes are colored according to community membership, and defined by the \cite{Gordon2016} atlas. The top row includes all the communities from the \cite{Gordon2016} atlas in the network, whereas the bottom row plots only the ``None'' and ``Visual'' communities in the network.  The \texttt{gordfmri} data is processed from the second scan of the first subject from the NYU test-retest resting-state fMRI data set}
  \label{fig:combined3dplot}
\end{center}
\end{figure}

We include an original visualization function in the \textbf{fabisearch} package that creates interactive, brain specific, 3-dimensional networks (Section~\ref{sec:networkestimation}). We use the networks computed in Section \ref{sec:eststatnet} as an example. We begin by plotting the network of the first stationary block of data (time points 1:35) using the threshold based method (Figure \ref{fig:combined3dplot}, top row).  Given the density in this network, we chose $\lambda= 0.50$, which results in a network that is very sparse and has approximately 200 edges. Since we are using the \cite{Gordon2016} atlas, the rest of the settings for \fct{net.3dplot} are the default values.
\begin{verbatim}
R> net.3dplot(thresh.net[[n]][[50]])
\end{verbatim}
Edges between the nodes are added iteratively using a \texttt{for} loop, so depending on the number of edges to be added, the \fct{net.3dplot} function may take some time to incorporate all elements of the network. In Figure \ref{fig:combined3dplot} (top row), most connections are concentrated in the ``Visual'' and ``None'' communities, with some dense connections in the ``DorsalAttn'', ``CinguloOperc'', and ``VentralAttn'' communities. Connections to nodes in other communities are also evident, however, they are not as densely interconnected. We also see the density of connections is more localized to the anatomical left hemisphere.  Furthermore, since most edges appear to be between the purple (``None'') and yellow (``Visual'') communities, we use \texttt{net.3dplot} to plot the 3-dimensional networks and allow us to focus on these specific communities (Figure \ref{fig:combined3dplot}, bottom row).  We specify the same node colors to maintain consistency between the plots (top and bottom rows of Figure \ref{fig:combined3dplot}), and add ROI labels.

\begin{verbatim}
R> communities = c("None", "Visual")
R> colors = c("#FFEB3B", "#673AB7")
R> net.3dplot(thresh.net[[n]][[50]], ROIs = communities,
+            colors = colors)
\end{verbatim}

In Figure \ref{fig:combined3dplot}, we show two angles of the brain network, from a caudal and lateral viewpoint, respectively.  The caudal view allows is to examine inter-hemisphere connections, while the lateral view allows us to see that most of the connections are in fact within the visual network.  In practice, however, the output of \fct{net.3dplot} opens a window for the 3-dimensional plot using the \textbf{rgl} library \citep{adler}. This plot is interactive in that it can be rotated by clicking and dragging the cursor on the plot. 

A major advantage of the \fct{net.3dplot} function in \textbf{fabisearch} is that we can also use different brain atlases to define the size and location of the ROIs, and estimate and plot the corresponding networks.  To the best of our knowledge, \textbf{fabisearch} is the first R package that has the ability to visualize an interactive, 3-dimensional, brain specific network that allows for various atlases and also for any manually uploaded coordinate inputs.  For example, in addition to the \texttt{gordatlas}, we also include the \texttt{AALatlas},  which is the 90 ROI Automated Anatomical Labelling (AAL) atlas as defined by \cite{tzourio} in \textbf{fabisearch}. In addition, the \texttt{AALfmri} data set in \textbf{fabisearch} is the same as the \texttt{gordfmri} data set, except it has 90 ROIs instead of 333 ROIs and it uses the AAL atlas parcellation instead of the \cite{Gordon2016} atlas parcellation.  For the \texttt{AALfmri} data, we use the same detected change points and re-estimate the network for the first stationary time segment, corresponding to time points 1:35 and save this as \texttt{AAL.net}.

\begin{verbatim}
R> data("AALfmri", package = "fabisearch")
\end{verbatim}

\begin{verbatim}
R> lambda = seq(0.01, 0.99, 0.01)
R> AAL.net = est.net(AALfmri[1:35,], lambda = lambda, rank = 6)
\end{verbatim}

\begin{figure}[h]
\begin{center}
  \includegraphics[width=0.35\linewidth]{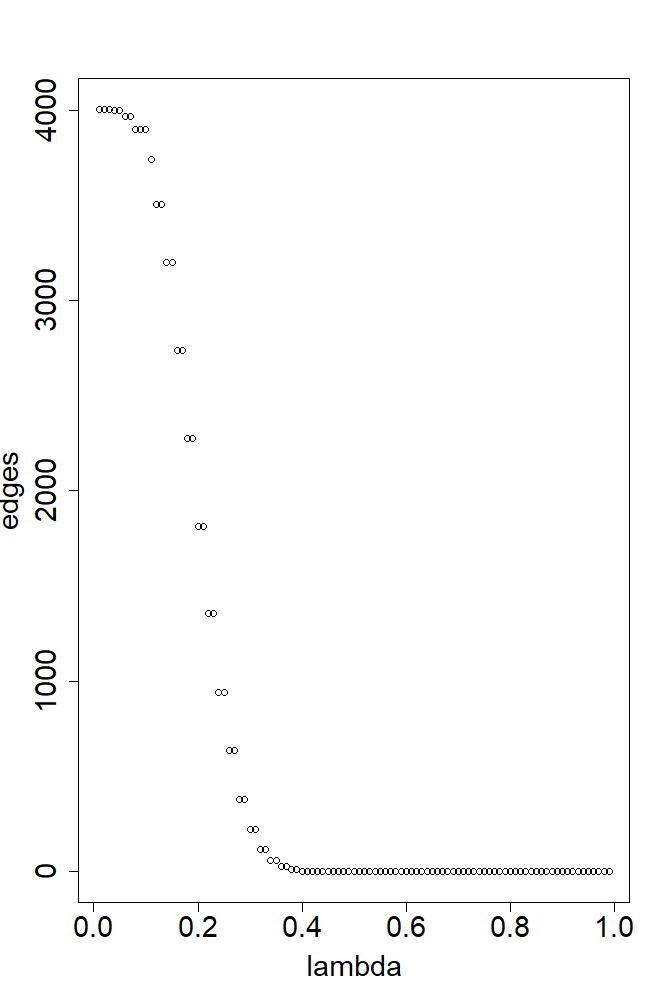}
  \caption{A plot of \texttt{lambda} values against the number of edges (or elements in the upper triangular part of the adjacency matrices) for the stationary network in the \texttt{AALfmri} data set for time points 1:35. The \texttt{AALfmri} data is processed from the second scan of the first subject from the NYU test-retest resting-state fMRI data set.}
  \label{fig:AALelbow}
\end{center}
\end{figure}

Again, we must find an appropriate cutoff for our thresholding of the consensus matrix so we explore a range of \texttt{lambda} values (Figure \ref{fig:AALelbow}).

\begin{verbatim}
R> edges = c()
R> for(i in 1:length(lambda)){
+   curr.net = AAL.net[[i]]
+   edges = c(edges, sum(curr.net[lower.tri(curr.net)]))
+ }
R> plot(lambda, edges, cex.lab = 1.8, cex.axis = 1.8)
\end{verbatim}

\begin{figure}[h]
\begin{center}
  \includegraphics[width=1\linewidth]{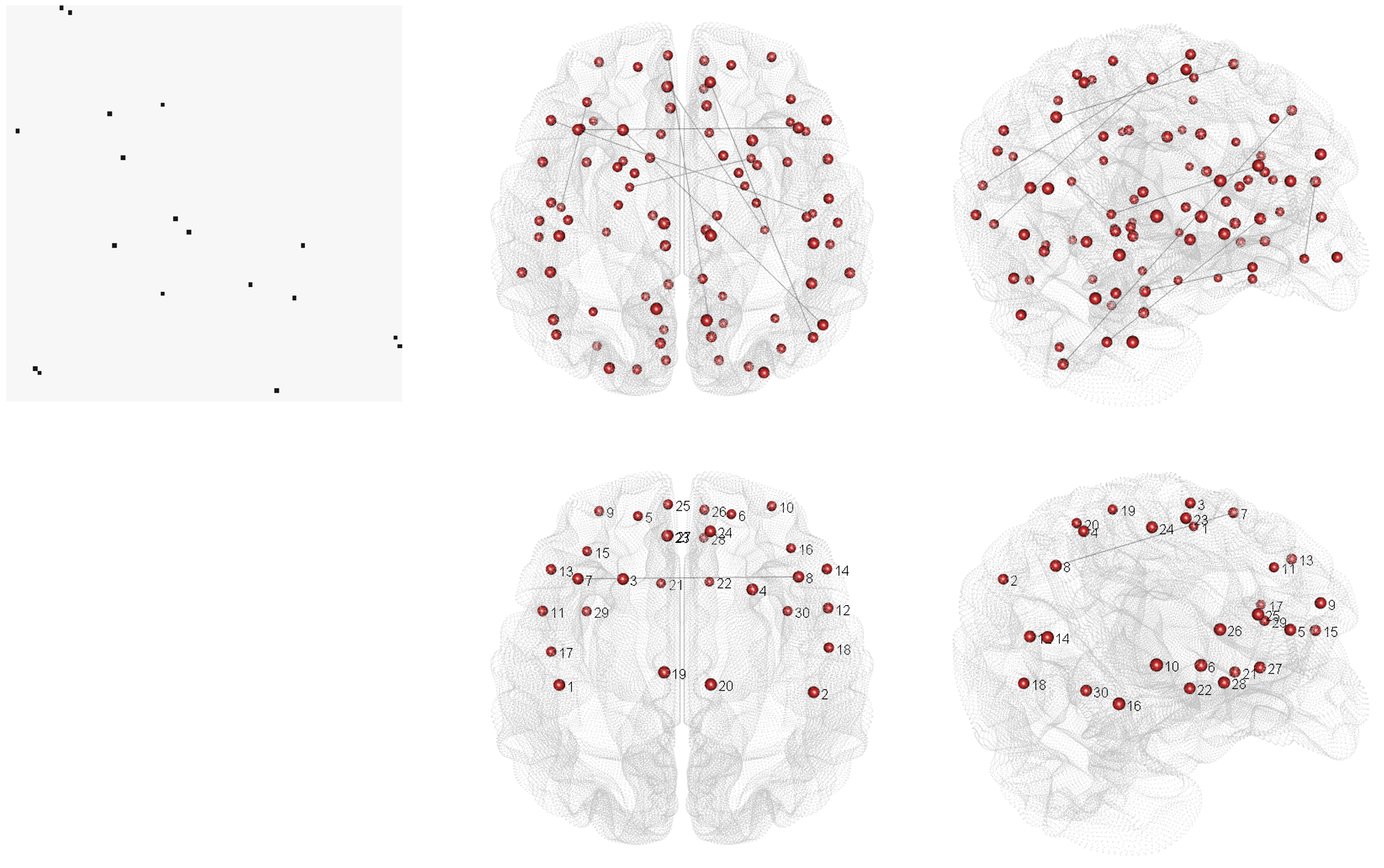}
  \caption{Visualizations of the estimated network from the first stationary block in the \texttt{AALfmri} data set, using the threshold based method where \texttt{lambda = 0.38}. The top row, includes an adjacency matrix, and a 3-dimensional brain plots at different angles using the entire AAL atlas, while the bottom row includes two plots are of the same network but with only relationships between the first 30 ROIs.  The \texttt{AALfmri} data is processed from the second scan of the first subject from the NYU test-retest resting-state fMRI data set. }
  \label{fig:AALplots}
\end{center}
\end{figure}

From Figure \ref{fig:AALelbow}, it is evident that a cutoff value equal to 0.38 for the threshold for the adjacency matrix, balances sparsity and interpretability.  We plot the adjacency matrix and the 3-dimensional brain with this cutoff value in Figure \ref{fig:AALplots} (top row) using:

\begin{verbatim}
R> heatmap(AAL.net[[38]], col = grey(c(0.97,0.1)), symm = TRUE,
+         Colv = NA, Rowv = NA, labRow = NA, labCol = NA)
R> net.3dplot(AAL.net[[38]], coordROIs = AALatlas)
\end{verbatim}

From Figure \ref{fig:AALplots} (top row), we observe that the network is relatively sparse. Compared to the first stationary network plotted for the \texttt{gordfmri} data set (Figure~\ref{fig:combined3dplot}, top row), the connections between nodes in the \texttt{AALfmri} first stationary network appear more diffuse. There is not a clear clustering structure, although it seems that there are more dense connections in nodes close to the mid-line of the brain. Additionally, the higher density of the left hemisphere appears to be preserved in this data set as well. We focus in on a subset of nodes by specifying the node numbers in the \fct{net.3dplot} under the \texttt{ROIs} parameter. In this case, we plot the networks in Figure \ref{fig:AALplots} (bottom row), where we narrow down the relationships to the first 30 ROIs (or nodes) using:

\begin{verbatim}
R> nodes = 1:30
R> net.3dplot(AAL.net[[38]], coordROIs = AALatlas, ROIs = nodes)
\end{verbatim}

Finally, as mentioned in Section~\ref{sec:visual}, \fct{net.3dplot} also provides capability for plotting interactive brain networks using any cortical atlas (which can be manually uploaded using specific coordinate inputs), and hence adjacency matrix, for defining nodes in the network.  Lastly, the visualization function is general; it can also be used
on data sets without any change points.

%-----------------------------------------------------------------------
\section{Financial Data Example} \label{sec:financial}
\subsection{Data summary and change point detection} \label{sec:FINdatacpd}

Although the \textbf{fabisearch} package was initially motivated, developed and tested for functional magnetic resonance imaging (fMRI) time series data, the methodology and package functionality could naturally extend to any multivariate high-dimensional change point detection problem. We now illustrate such an example.

Another popular area of application for change point detection and network estimation is financial data.  Here the objective is to find clusters of stocks with strong temporal dependence. In portfolio management, this is useful in order to spread risk or variability of returns over time by seeking to minimize dependence between selected components of the portfolio. Naturally, these dependencies change over time as the general market milieu changes and the companies themselves adapt. As such, knowing when these changes occur is of high value as it might signal a time when portfolio allocations should be re-evaluated.  To this end, using periodic returns from individual stocks over time, we apply the FaBiSearch methodology as described in Section \ref{sec:models} to the Standard and Poor's 500 (S\&P 500) index (\texttt{\url{https://www.spglobal.com/spdji/en/indices/equity/sp-500}}), which is a market capitalization weighted index of 500 large companies listed on United States stock exchanges. We first obtained a list of the 505 ticker symbols (which is greater than the number of companies because some companies have multiple stock tickers), and then acquired each stock's publicly available, daily historical price from Yahoo Finance (\texttt{\url{https://finance.yahoo.com/}}) for dates 2018/01/01 to 2021/03/31. Dates in this range when the stock market was closed were removed from the data set. To clean the data, we first removed any companies which were not publicly listed for the entirety of our selected time period, thus removing any variables which have missing data. Next, we standardized the data by calculating the daily $log$ returns over the aforementioned period. 
%Then, we scaled the $\sigma$ by multiplying each entry by the same factor, in this case 3, to accentuate the absolute difference between data points. 
To allow for the use of non-negative matrix factorization (NMF), we added 100 to each entry to increase the mean of each time series to 100. The fully processed data set has 815 rows/dates, 499 variables/stocks, and is available as the \texttt{logSP500} data set in the \textbf{fabisearch} R package.

\begin{figure}[h]
\begin{center}
  \includegraphics[width=1\linewidth]{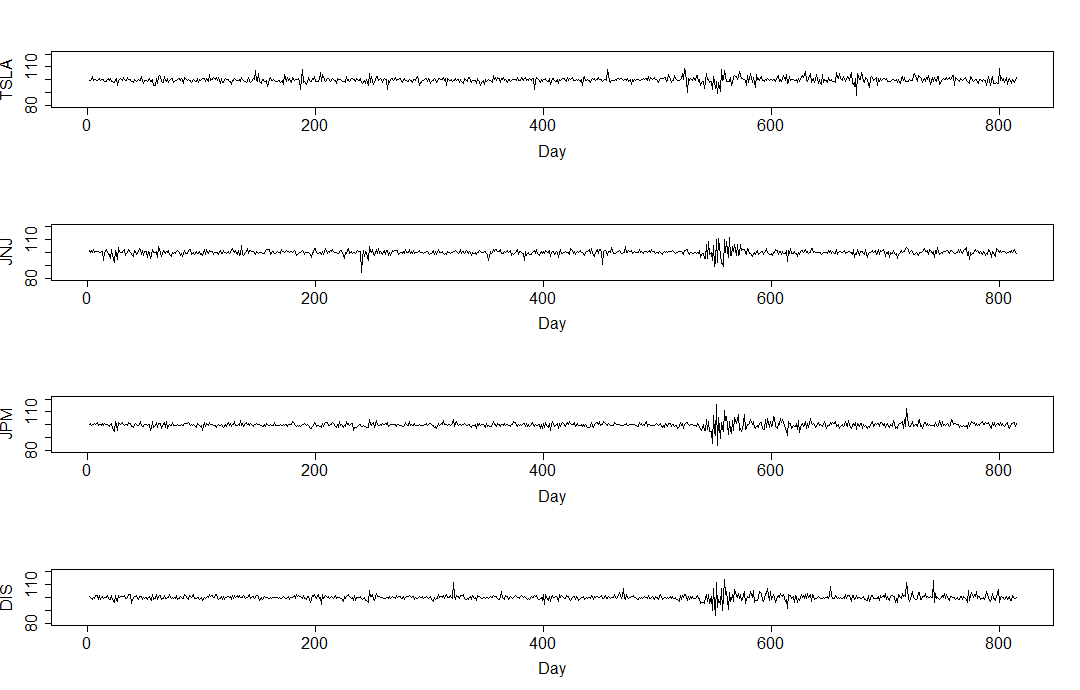}
  \caption{Plots of the daily adjusted log-returns (y-axis) for companies TSLA, JNJ, JPM, and DIS in the S\&P 500 index for dates 2018/01/01 to 2021/03/31 (x-axis), from the \texttt{logSP500} data set in the \textbf{fabisearch} R package.}
  \label{fig:SP500ts4}
\end{center}
\end{figure}

To begin, we inspect the data by calculating summary statistics and plotting individual time series (Figure \ref{fig:SP500ts4}) for four prominent companies included in this index, namely Tesla Inc (TSLA), Johnson \& Johnson (JNJ), JPMorgan Chase \& Co. (JPM), and Walt Disney Company (DIS). 

\begin{verbatim}
R> data("logSP500", package = "fabisearch")
R> companies = c("TSLA", "JNJ", "JPM", "DIS")
R> columns = colnames(logSP500) %in% companies
R> print(logSP500[1:10, columns])
                 JNJ       JPM      TSLA       DIS
2018-01-03 101.35288 100.05630  99.35006 100.37610
2018-01-04  99.95933 100.83552  99.44736  99.89045
2018-01-05 101.16552  99.33067 100.16846  99.37655
2018-01-08 100.15435 100.10079 102.87447  98.45363
2018-01-09 102.25866 100.44912  99.45760  99.86150
2018-01-10  99.79798 101.02189 100.02511  99.49626
2018-01-11 100.79597 100.47644 100.32494 101.35214
2018-01-12 100.94112 101.55029  99.60574 101.29637
2018-01-16 101.06352  99.61123 100.42376  98.29807
2018-01-17 100.08850 100.57912 100.88551 101.11678
R> summary(logSP500[, columns])
      JNJ              JPM              TSLA             DIS        
 Min.   : 84.58   Min.   : 84.19   Min.   : 88.13   Min.   : 85.65  
 1st Qu.: 99.14   1st Qu.: 99.15   1st Qu.: 98.95   1st Qu.: 99.11  
 Median :100.07   Median :100.00   Median : 99.96   Median : 99.98  
 Mean   :100.00   Mean   :100.00   Mean   :100.00   Mean   :100.00  
 3rd Qu.:100.96   3rd Qu.:100.87   3rd Qu.:100.97   3rd Qu.:100.87  
 Max.   :111.16   Max.   :116.07   Max.   :108.86   Max.   :113.76  
 
R> par(mfrow=c(4,1))
R> for(i in 1:4){
+   plot(logSP500[,colnames(logSP500) == companies[i]], 
+        type = "l", cex.lab = 1.5, cex.axis = 1.5, xlab = "Day", 
+        ylab = companies[i], ylim = c(80, 120))
+ }
\end{verbatim}

We continue our analysis by applying FaBiSearch to the data in order to detect change points in the network (or clustering) structure of this multivariate high-dimensional time series. We utilize the \fct{detect.cps} function (as described in Section \ref{sec:datacpd}), but because of the greater number of time points in this data set we increase \texttt{mindist}, the minimum distance between candidate change points, in the \texttt{detect.cps} function, to 100. Additionally, due to the larger number of nodes, we increase \texttt{nreps} to 150 to increase the power of the statistical inference test on the candidate change points. Again, we assume the appropriate rank to use for this data set is unknown a priori, hence we use \fct{opt.rank} to determine the appropriate rank. This is saved as \texttt{rank}, and is then fed into the \fct{detect.cps} function. All other arguments for \fct{detect.cps} remain at their default values. The final output of the change point detection procedure are saved as the \texttt{SP500out} object.

\begin{verbatim}
R> set.seed(54321)
R> rank = opt.rank(logSP500)
R> SP500out = detect.cps(logSP500, alpha = 0.05, mindist = 100, 
+                       rank = rank, nreps = 150)
\end{verbatim}

The corresponding output is, again, a list with three components. Interestingly, the optimal rank for this data set is lower than that for the fMRI data, indicating a simpler clustering structure with high dependence amongst components. This is quite plausible given the index is made up from the same asset class (equities), focuses on the the United States market exclusively, and is made up of large-capitalization companies only.

\begin{verbatim}
R> SP500out
$rank
[1] 4

$change_points
    T stat_test
1 115      TRUE
2 215     FALSE
3 315     FALSE
4 460      TRUE
5 560      TRUE
6 660      TRUE

$compute_time
Time difference of 48.32602 mins
\end{verbatim}

Four change points at time points $t = 115, 460, 560, 660$ were detected and saved in \texttt{finalcpt}.

\begin{verbatim}
R> finalcpt = SP500out$change_points$T[SP500out$change_points$stat_test]
R> finalcpt
[1] 115 460 560 660
\end{verbatim}

We now relate these change points to the actual dates, which are stored in \texttt{logSP500} as the rownames.

\begin{verbatim}
R> rownames(logSP500[finalcpt,])
[1] "2018-06-18" "2019-10-30" "2020-03-25" "2020-08-17"
\end{verbatim}

The four change points correspond to five unique, macro-level environments which dictate the underlying clustering and network structure of the stocks in the index. In the first segment, there is relatively steady growth of the index during 2018. The first change point characterizes a switch to a higher volatility market environment, in which the S\&P 500 dropped approximately $18\%$ as fear of trade tensions between the US and China began to mount. The index continued to experience turbulence as the S\&P 500 attempted to hit new all time highs. The point where the index lifted beyond this ceiling corresponds to the second change point. Afterwards, the recovery was quite strong until in early 2020 when the second major hit of volatility occurred, coinciding with the beginning of the COVID-19 pandemic. The end of this volatility is marked by the third change point, after which the index rebounded sharply, as the index continued to drive higher from the pandemic related lows. During this time, many tech companies saw accelerated growth from increased web traffic and demand for technology during stay at home mandates, while the retail, services, and travel industries remained stagnant. Near the end of 2020 marks the final change point, which preceded a slight change in environment as the market became more selective about valuations and which stocks could justify their growth. This also coincides with the market testing previous all time highs, which further added to the general hesitancy about market levels.

\subsection{Estimating stationary networks and visualization} \label{sec:FINeststatnet}

We continue our analysis of this data set by estimating the stationary networks for the five segments. Given the optimal rank calculated for this data set, we specify \texttt{rank = 4} in the \fct{est.net} function. We also increase the number of runs to $n_{runs}=100$ to improve the stability of the calculated networks. Again, we specify the \texttt{changepoints} parameter using the \texttt{finalcpt} vector. We save the output as \texttt{SP500net}.

\begin{verbatim}
R> SP500net = est.net(logSP500, rank = rank, 
+                    lambda = rank, nruns = 50, changepoints = finalcpt)
\end{verbatim}

Given the large number of variables in this data set, we focus on the 15 largest companies in the S\&P 500 index by market capitalization at the time of writing. We define \texttt{tickers}, a vector of string values which contains the ticker symbols for these companies, and narrow down the networks to the relationship between these 15 companies.

\begin{verbatim}
R> tickers = c("AAPL", "MSFT", "AMZN", "GOOG", "FB", "TSLA", "BRK-B", 
+             "JPM", "JNJ", "NVDA", "UNH", "V", "HD", "PG", "DIS")
\end{verbatim}

The cluster based approach works well for a smaller number of nodes, as the number of edges is relatively manageable. We use the \textbf{igraph} package to plot the graphs.

\begin{verbatim}
R> library(igraph)
R> top.15 = colnames(logSP500) %in% tickers
R> par(mfrow=c(2,2))
R> for(i in 1:length(SP500net)){
+     G = graph_from_adjacency_matrix(SP500net[[i]][top.15,top.15],
+          mode = "undirected", diag = FALSE)
+     V(G)$color = palette()[components(G)$membership]
+     plot(G, layout=layout_with_kk, vertex.size=20,
+          vertex.color="white", vertex.frame.color = "white",
+          vertex.label.color = V(G)$color, vertex.label.family =
+          "Helvetica", asp = 0.5)
+     Sys.sleep(1)
+ }
\end{verbatim}

\begin{figure}[ht]
\begin{center}
  \includegraphics[width=0.75\linewidth]{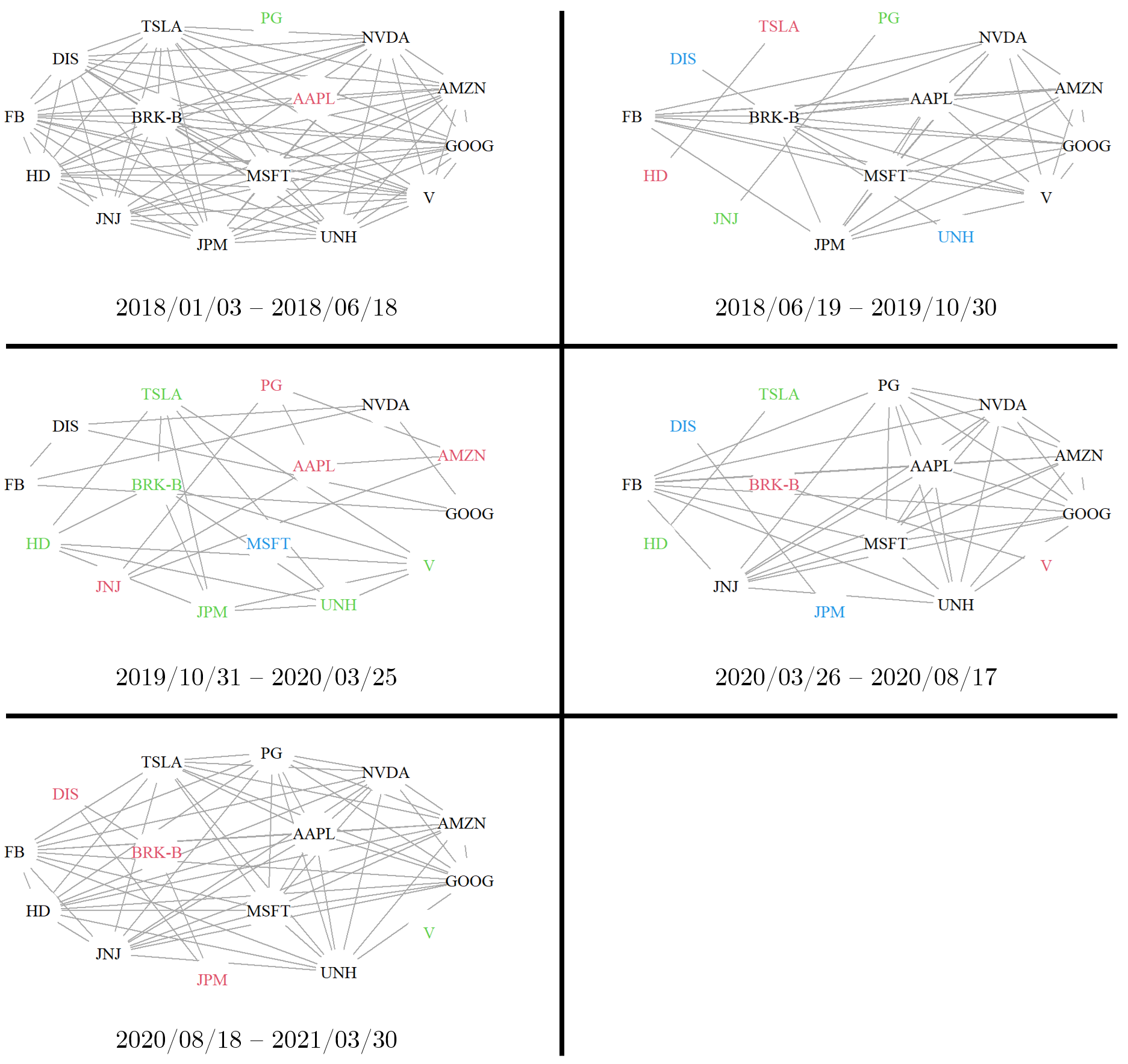}
  \caption{The five stationary networks of the top 15 largest S\&P 500 companies by market capitalization from 2018/01/03 to 2021/03/30 estimated using the function \fct{est.net}. Each network is made up of up to 4 clusters, calculated using a rank of 4 for 50 runs, and is labelled with corresponding date ranges below. Cluster membership is denoted by the vertex label colors.}
  \label{fig:SP500nets}
\end{center}
\end{figure}

Figure \ref{fig:SP500nets} shows the stationary networks between each pair of change points. There is evidence of a changing clustering and network structure between stationary segments for the selected 15 companies. At a broad level, the companies became more densely connected during periods where the index experienced steady, sustainable growth (the first and last stationary segments). For the first network (Figure \ref{fig:SP500nets}, top left), the companies are separated into three unique clusters. We see that the big tech companies, outside of Apple (AAPL) are contained in the black cluster. After the first change point however, and as evident in the next network (Figure \ref{fig:SP500nets}, top right), the edges and cluster memberships change. The network in general becomes more sparse, with companies separating into four clusters. The core big tech companies remain in the black cluster, however Tesla (TSLA) is displaced by AAPL. Disney (DIS), Johnson \& Johnson (JNJ), and United Health Group (UNH) are also separated from the black cluster. After the second change point, the density of edges decreases again, and the individual associations are greatly changed (Figure \ref{fig:SP500nets}, middle left). For example, the big tech group is dissolved, as Microsoft (MSFT), Amazon (AMZN), and AAPL dissociate from the black cluster. Additionally, MSFT is separate in its own cluster is a behaviour unique only to this stationary segment. Both of the aforementioned cases are examples where the associations are out of the norm, or atypical. Furthermore, we see that the general network structure is much more dispersed, with more smaller clusters dominating the network. This likely due to the extreme market volatility, wherein companies that typically are not associated with one another become clustered together in the estimated network. After the third change point, the next stationary segment (Figure \ref{fig:SP500nets}, middle right) corresponds to the sharp rebound in the index and we see a high concentration of companies into a single cluster. In this segment, there are four clusters, however the network is particularly dominated by the black cluster which includes 9 of the 15 companies. The final stationary segment (Figure \ref{fig:SP500nets}, bottom left) occurs after the fourth and final change point, and shows some similarity in structure to the first stationary segment. Here, there are again three clusters albeit the network is heavily dominated by the black cluster, which includes 11 of the 15 companies.

\section{Discussion and Conclusion} \label{sec:summary}
The \textbf{fabisearch} R package is a comprehensive software implementation of the FaBiSearch methodology and can also be used to visualize interactive 3-dimensional brain specific networks. We utilize the \textbf{NMF} package \citep{Gaujoux2010} as a starting point for our implementation of non-negative matrix factorization (NMF) in \textbf{fabisearch}, parallelizing the code to maximize computational efficiency. Specifically, NMF fitments and the refitting procedure in FaBiSearch are parallelized which greatly reduces computational time in multi-threaded machines. 

It is important to note that, in practice, NMF, and consequently FaBiSearch, is sensitive to the input data set $\boldsymbol{Y}$. NMF can be considered most simply as an algorithm which seeks to minimize loss between observed data and the model. A stepwise approach is used to minimize the objective function, and NMF stops when the difference in the objective function between subsequent steps is negligible. Consequently, NMF is sensitive to scale and may stop prematurely if the absolute difference between input data points is small. As such, we recommend re-scaling data to ensure that the standard deviation, $\sigma \geq 2$, if the mean is approximately equal to $100$. Naturally, as $\mu$ increases, $\sigma$ should increase proportionately and vice versa. If starting with data where $\mu = 0$, it can be scaled by multiplying each entry in $\boldsymbol{Y}$ by some factor. This is what was performed to scale up $\sigma$ in the log returns for the \texttt{logSP500} data set in Section~\ref{sec:financial}. Then, 100 was added to each data entry to ensure that no negative values remained (new $\mu = 100$) and to make it suitable for use with NMF.

The computational time for functions \fct{detect.cps} and \fct{est.net} is sensitive to the size of the input data, and the \texttt{rank}, \texttt{nruns}, \texttt{nreps}, and \texttt{algtype} arguments chosen. Naturally, as \texttt{rank}, \texttt{nruns}, and \texttt{nreps} increase, so does the computational time. However, for a data set on the order of 100-200 nodes, and 200-400 time points, most uses of \fct{detect.cps} are on the order of minutes to possibly a couple hours on consumer grade machines, which is reasonable, given the dimension of the data. Further, the graph estimating function, \fct{est.net}, for the same size of data set executes on the order of seconds to minutes.

NMF, and thus FaBiSearch, is also sensitive to the user specified input parameters. From the arguments in \fct{detect.cps} and \fct{est.net}, inputs \texttt{rank}, \texttt{nruns}, \texttt{mindist}, \texttt{nreps}, and \texttt{alpha} are the most important. Typically, \texttt{rank} can be accurately assessed using the automated optimization procedure by specifying \texttt{rank = "optimal"}. In general however, we recommend to overestimating \texttt{rank} if specifying it manually. When specifying \texttt{nruns} it is important to be mindful of overfitting, and for most applications a value between 20-250 is appropriate. The specific value which is most appropriate for a data set depends on the number of variables or columns, but the number of samples or time points is also influential as well. The value of \texttt{mindist} should be chosen such that the sample of $\boldsymbol{Y}$ is sufficiently large to ensure a stable NMF fitment. Lastly, \texttt{nreps} and \texttt{alpha} are parameters more well understood in statistics; general speaking, a larger \texttt{nreps}, and a more stringent and therefore lower value of \texttt{alpha} are more favourable.

To illustrate how the values of \texttt{rank} and \texttt{nruns} affects the accuracy of FaBiSearch, we provide a brief sensitivity analysis on a simulated data set. This data set is effectively the same as the \texttt{sim2} data set in the \textbf{fabisearch} package, although the number of variables has been increased to $p = 200$.  More specifically, data are generated from the multivariate Gaussian distribution $\mathcal{N}(0, \Sigma)$, with the following structures on $\Sigma$:
\begin{equation*}
\Sigma_{ij} = \begin{cases}
0.75, & \mbox{if } i\neq j \mbox{ and } i, j \mbox{ are in the same cluster};\\
 1, & \mbox{if } i = j;\\
0.20, & \mbox{otherwise},
 \end{cases}
\end{equation*}

\begin{figure}[ht]
\begin{center}
  \includegraphics[width=0.9\linewidth]{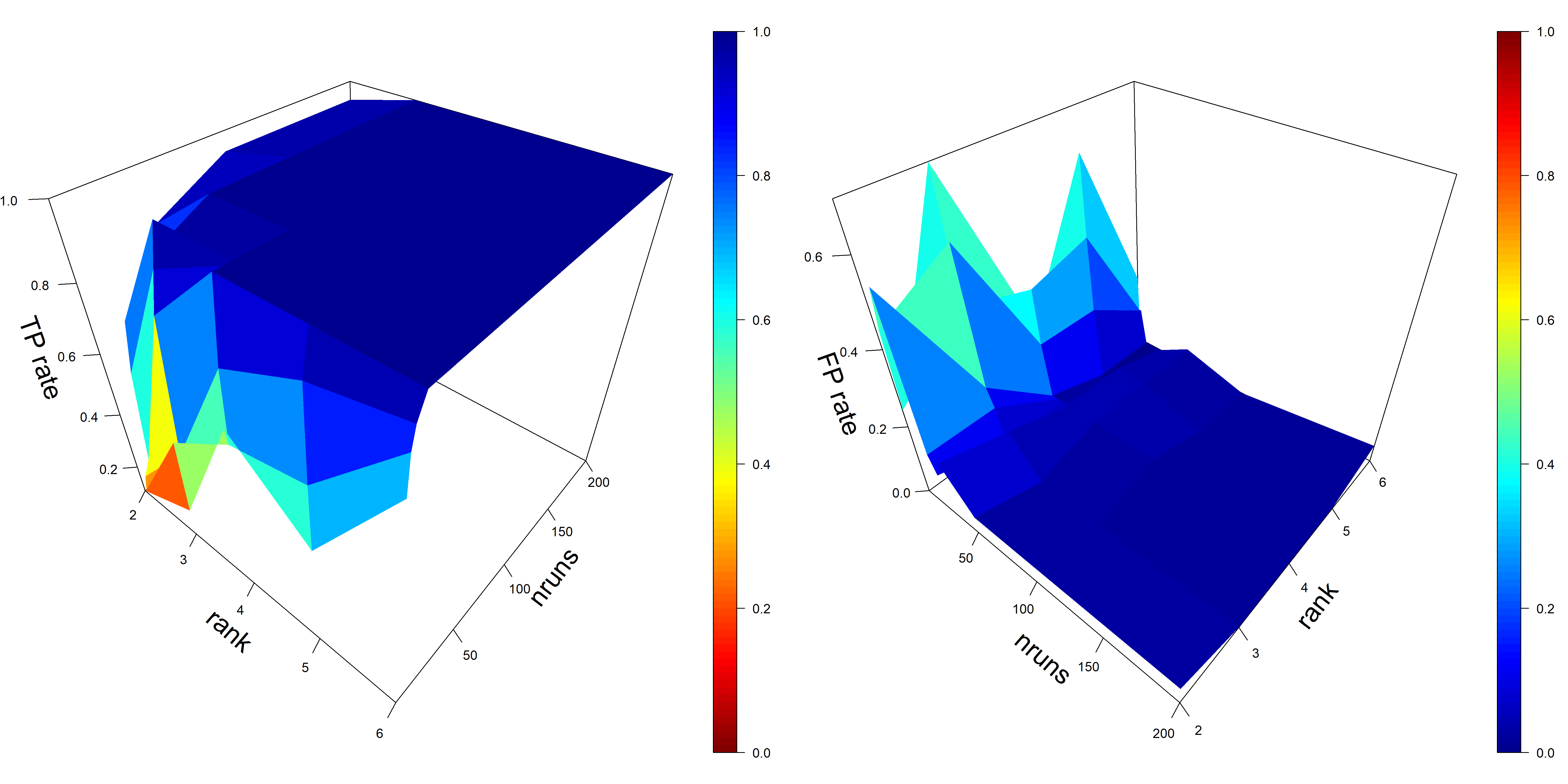}
  \caption{A sensitivity analysis on different combinations of factorization rank, $r$ and the number of runs, $n_{runs}$ using a simulated data set.  The visualization occurs from two perspective plots: true positive (TP) and false positive (FP) rates are shown on the left hand side and right hand side, respectively. Note the x and y axes are flipped between the two plots to more clearly see the sharp changes to accuracy before it plateaus. Color scales are also flipped, however blue is more favourable (higher TP, lower FP) and red is less favourable (lower TP, higher FP) in both plots.}
  \label{fig:sensanalysis}
\end{center}
\end{figure}

The length of the time series remains as $T = 200$, where the first half is characterized $\Sigma$. At $t = 100$, there is a change point where the node labels are reshuffled randomly amongst the two clusters in $\Sigma$. This is then repeated to create a sample of 20 unique instances of this data set. 50 combinations of \texttt{rank} and \texttt{nruns} were then tested, using values of \{2, 3, 4, 5, 6\}, and \{2, 3, 5, 10, 15, 20, 25, 50, 100, 200\}, respectively. For each of these combinations, true positive (TP) and false positive (FP) rates were calculated for the 20 samples. We define a TP as a change point found within 5 time points ($t = 100 \pm 5$), and a FP as a change point found outside of these bounds ($t \neq 100 \pm 5$). The results are shown in Figure \ref{fig:sensanalysis} as two perspective plots for the TP and FP rates. Note the $x$ and $y$ axes are flipped between the two plots to more clearly see the sharp changes to accuracy before it plateaus. Color scales are also flipped, however blue is more favourable (higher TP, lower FP) and red is less favourable (lower TP, higher FP) in both plots. For both TP and FP rates, there is a steep improvement by increasing \texttt{nruns} from 2 to 25, and \texttt{nruns} in general plays a key role in the accuracy of \fct{detect.cps}. After this, increasing \texttt{nruns} shows only marginal improvements and accuracy effectively plateaus. This falls in line with expectations, as only the best and therefore lowest loss NMF fitment is used when searching for change points. Naturally, as \texttt{nruns} increases the probability of finding a good fitment similarly increases. Once enough random starts have been attempted however, it is unlikely that a better fitment can be found. Additionally, while \texttt{rank} is also important, it does not contribute as greatly to accuracy in this data set. Increasing \texttt{rank} at lower values of \texttt{nruns} yields marginally better results. However, this data set has a relatively simple clustering structure and so for data sets where it is more complicated \texttt{rank} should play a greater role. We also compare our novel binary search based segmentation method to the more typical, binary segmentation in the Appendix. We show, through a challenging simulation study, that this new method is both more accurate and efficient. Lastly, while our model setup does not accomodate for changes in rank, in general, the experimental results of \cite{ondrus2021factorized} suggest that as long as the rank is greater than or equal to the largest number of clusters in any one time segment, the accuracy of the method plateaus or improves marginally. This type of behaviour is quite typical for NMF, and clustering algorithms more generally.

In this paper, we delineate the four main functions included in the \textbf{fabisearch} package, and apply them to two distinct multivariate high-dimensional time series (a fMRI study, and the S\&P 500 stock index). Further, we show how the functionality of \fct{detect.cps} and \fct{est.net}, for change point detection and network estimation respectively, generalizes to any multivariate time series problem. We also provide specific functionality to plot interactive 3-dimensional brain networks using the \fct{net.3dplot} function, and we emphasize that it can be used with any cortical atlas (and for any manually uploaded coordinate inputs), and hence adjacency matrix, for defining nodes in the network.  It can also be used on data sets without any change points.  This, in combination with the ability to select sub-networks and color nodes by community membership, provides great flexibility for displaying elegant 3-dimensional brain networks. Finally, we provide a brief sensitivity analysis on a simulated data set to characterize how values of \texttt{rank} and \texttt{nruns} affect NMF and change point detection accuracy using \fct{detect.cps()}.

\section*{Computational Details} \label{sec:compdetails}

The results in this paper were obtained using
R~4.1.3 with the \textbf{fabisearch}~0.0.4.4 package. R itself and the \textbf{fabisearch} package are available from the Comprehensive R Archive Network (CRAN) at \url{https://cran.r-project.org} and \url{https://cran.r-project.org/package=fabisearch}, respectively. Computations performed using two Intel Platinum 8160F Skylake at 2.1Ghz (48 cores total) and 187GB of memory.

\section*{Acknowledgments}

This research was enabled in part by support provided by WestGrid (www.westgrid.ca) and Compute Canada (www.computecanada.ca). The first author was supported by the Natural Sciences and Engineering Research Council of Canada (NSERC) Alexander Graham Bell Master's Scholarship, the Alberta Innovates Graduate Student Scholarships (Alberta Innovates, Alberta Advanced Education), and the Richard B. Stein Studentship (Neuroscience and Mental Health Institute, University of Alberta). The second author was supported by the Natural Sciences and Engineering Research Council (Canada) grant RGPIN-2018-06638 and the Xerox Faculty Fellowship (Alberta School of Business).

\newpage
\bibliographystyle{abbrvnat}
\bibliography{fabisearch.bib}

\newpage
\appendix
\section*{Appendix}
\subsection*{Comparison of binary search and binary segmentation}
\label{app:appendix}

In this experiment, we compare FaBiSearch and the binary search based change point detection method to the more typical binary segmentation method. The binary segmentation method is a similar methodology but switches the binary search to a \texttt{for} loop, which akin to a brute force search, or grid search over all possible change points. In this comparison, the settings for the methods are as follows: \texttt{rank = 3, mindist = 35, nruns = 50, nreps = 100, alpha = NULL, algtype = "brunet", testtype = "t-test"}.

\noindent The first code chunk is the result for the typical binary segmentation algorithm:

\begin{verbatim}
$rank
[1] 3
$change_points
    T    stat_test
1  41 1.0000000000
2 100 0.0000309575
3 156 0.6129467750
$compute_time
Time difference of 36.6964 mins
\end{verbatim}

\noindent The following code chunk is for the binary search based algorithm, which is used in our FaBiSearch methodology:

\begin{verbatim}
$rank
[1] 3
$change_points
    T    stat_test
1  52 1.000000e+00
2  99 3.628777e-07
3 165 1.000000e+00
$compute_time
Time difference of 5.119113 mins
\end{verbatim}

The true change point is at time point $t = 100$. Both algorithms obtain results that is close to this point.  However, the main difference is that the binrary segmentation method takes 36.70 minutes, while the binary search based algorithm takes 5.12 minutes, which represents a significant increase in computational efficiency. 

To compare the segmentation methods in a multiple change point situation, we also consider Simulation 3 from \citep{ondrus2021factorized}, which has two change points at time points $t=100$ and $t=200$. Table \ref{tab:app} shows the results of 100 repetitions of the simulation. For both the binary segmentation and the binary search, we used the following settings: \texttt{rank = 3, mindist = 50, nruns = 40, nreps = 100, alpha = 0.01, algtype = "brunet", testtype = "t-test"}.

\begin{table}[H]
    \begin{center}
        \begin{tabular}{cccccccc}
            \addlinespace[0.15cm]
            Search type & $t^*$ & TP 1 & FP 1 & TP 10 & FP 10 & Haus. dist. & Compute mins. \\ \hline\hline
            Binary segmentation & 100 & 0.25 & 1.12 & 0.47 & 0.61 & 0.3352 & 422.79 \\
            ~ & 200 & 0.21 & ~ & 0.50 & ~ & ~ & ~ \\ \hline
            Binary search & 100 & 0.81 & 0.39 & 0.99 & 0.01 & 0.0074 & 23.33 \\
            ~ & 200 & 0.80 & ~ & 1 & ~ & ~ & ~ \\ \hline
        \end{tabular}
    \end{center}
    \caption{A comparison of the binary segmentation and the binary search methods using simulated data. The true change points are denoted by $t^*$. TP 1 and TP 10 are true positives within +/- 1 and 10 time points, respectively. FP 1 and FP 10 are false positives outside +/- 1 and 10 time points, respectively. Haus. dist. and compute mins. are the average Hausdorff distance and computational time, respectively.}
    \label{tab:app}
\end{table}

\noindent Our results show that the binary search in FaBiSearch is orders of magnitude both more accurate (higher TP rates, lower FP rate and lowerHausdorff distance), as well as considerably more computaitonally efficient (lower compute time). 

\end{document}